\documentclass[final]{ws-ijmpa}
\def\draftnote{}
\usepackage{amsmath,amssymb}
\usepackage{epsfig}
\usepackage{hyperref}

\def\as{\ensuremath{\alpha_{s}}}
\def\ay{\ensuremath{\alpha_{y}}}
\def\meps{\ensuremath{m_{\varepsilon}}}
\def\Lmes{L}
\def\LME{LME}
\newcommand\myunderline[1]{\underline{\vphantom{G_{\mu}} #1}}
\def\scale{\ensuremath{\mu}}

\def\DR{\ensuremath{\overline{\mathrm{DR}}}}
\def\MSbar{\ensuremath{\overline{\mathrm{MS}}}}
\def\MS{\ensuremath{{\mathrm{MS}}}}
\def\MOM{\ensuremath{{\mathrm{MOM}}}}

\def\mbdr{\ensuremath{m_b^{\DR}}}
\def\mbms{\ensuremath{m_b^{\MSbar}}}

\def\GeV{\mbox{GeV}}
\def\TeV{\mbox{TeV}}
\def\Mgut{\ensuremath{M_{\mathrm{GUT}}}}
\def\Mew{\ensuremath{M_Z}}

\def\L{\ensuremath{ {\mathcal L} } }
\def\Tr{{\rm Tr}}

\newcommand{\dzeta}[2]{ \ensuremath{ \delta \zeta_{#1}^{(#2)} } }

\newcommand{\figbox}[1]{\raisebox{1.1cm}{#1}}
\begin{document}
\markboth{A.V.~Bednyakov}
{Running mass of the $b$-quark in QCD and SUSY QCD}

%
\catchline{}{}{}{}{}
%

\title{RUNNING MASS OF THE $B$-QUARK \\ IN QCD AND SUSY QCD}
\author{A.V.~BEDNYAKOV}
\address{BLTP, Joint Institute for Nuclear Research, Dubna, Russia\\
bednya@theor.jinr.ru}

\date{\today}
\maketitle

\begin{abstract}
	The running mass of the $b$-quark defined in \DR-scheme 
	is one of the important parameters of SUSY QCD.
	To find its value, it should be related to some
	known experimental input. 
	In this paper, the $b$-quark running mass 
	defined in nonsupersymmetric QCD is chosen
	for determination of the corresponding parameter in
	SUSY QCD.
	The relation between these two quantities 
	is found by considering five-flavor QCD 
	as an effective theory obtained from its supersymmetric extension.
	A numerical analysis of the calculated two-loop relation 
	and its impact on the MSSM spectrum is discussed.
	Since for nonsupersymmetric models 
	\MSbar-scheme is more natural than \DR,  
	we also propose a new procedure that allows one
	to calculate relations between \MSbar- and \DR-parameters. 
	Unphysical $\varepsilon$-scalars that give
	rise to the difference between the above-mentioned schemes 
	are assumed to be heavy and decoupled 
	in the same way as physical degrees of freedom.
	By means of this method
	it is possible to ``catch two rabbits'', i.e., 
	decouple heavy particles and turn from \DR\ to \MSbar, at the same time.
	An explicit two-loop example of $\DR\to\MSbar$ transition 
	is given in the context of QCD.		
	The advantages and disadvantages of the method are 
	briefly discussed.
\keywords{QCD; MSSM; $b$-quark}
\end{abstract}

\ccode{PACS numbers: 12.38.-t, 12.38.Bx, 12.60Jv, 14.65Fy}

\section{Introduction}

It is commonly believed that the Standard Model (SM) is not the
ultimate theory of particle physics. Among other deficiencies of the SM
there is so-called fine tuning problem which arises from quadratic
dependence of the Higgs mass on the new physics scale.

	A popular extension of the SM that cures this problem is the Minimal
	Supersymmetric Standard Model (MSSM). The construction of the 
	CERN Large Hadron Collider (LHC) has led to many increasingly precise
	calculations of sparticle production and decay processes. 

	An important ingredient of the model is the SUSY QCD sector.
	In most of the processes with color particles radiative corrections
	from the strong interactions give a dominant contribution.
	Loop corrections to tree-level processes in SUSY QCD
	are usually expressed in terms of 
	running parameters defined in so-called \DR-scheme. It is
	an analog of \MSbar\ renormalization scheme based 
	on Dimensional Reduction (DRED)\cite{Siegel:1979wq}.	

	The running mass %
	\mbdr{} of the $b$-quark is one
	of the important parameters of SUSY QCD. The value of 
	\mbdr{} at a scale $\scale$ should be obtained from 
	some experimental input. However, for the $b$-quark
	it is very hard to find or even define such an input. 
	The pole mass $M_b$ being very well defined 
	in a finite order of perturbative QCD\cite{Tarrach:1980up,Kronfeld:1998di} 	
	suffers from renormalon
	ambiguity\cite{Bigi:1994em,Beneke:1994sw} that gives rise
	to $\Lambda_{QCD}/m_b \sim 10~\%$ uncertainty in its determination.
	There is another issue in using the pole mass for 
	determination of \mbdr{}. The relation
	between these two quantities exhibits a logarithmic
	dependence on all mass scales of SUSY QCD. This is a typical,
	nondecoupling, property of minimal 
	renormalization 
	schemes. In our problem we have very different scales, i.e., 
	$m_b \ll M_{SUSY}$. Consequently, 
	one cannot make all the logarithms small by some choice of the
	renormalization scale $\scale$, thus leading to 
	inaccurate perturbative prediction for  \mbdr.  

	A convenient quantity to use for extraction of \mbdr{}
	is the $b$-quark running mass 
	$m^{\MSbar}_b$ defined
	in the five-flavor QCD renormalized  
	in \MSbar-scheme\cite{Baer:2002ek}.
	The value of the running mass at the scale which is
	equal to itself is known from PDG\cite{Yao:2006px}, $\mbms (\mbms) = 4.20\pm0.07~\GeV$ . 
	
	In this paper, we calculate an explicit two-loop relation between
	$\mbms$ and $\mbdr$ by the so-called 
	matching procedure (see, e.g., Refs.~\refcite{Rothstein:2003mp} and \refcite{Georgi:1994qn}).

	In Section~\ref{sec:decoupling_framework}, our theoretical framework is described. 
	Renormalizable QCD is considered as an effective theory 
	that can be obtained from a more fundamental\footnote{
	In what follows we use adjectives ``full'', ''fundamental'' and ''high-energy''
	as synonyms to distinguish a more fundamental theory from the effective one}	
	one 
	by decoupling of heavy particles. 
	Decoupling of heavy degrees of freedom 
	manifests itself in relations between parameters
	of fundamental and effective theories.	
	Intuitively, in the energy region far below the 
	corresponding threshold contribution of virtual heavy loops 
	to the light particle effective action
	can be approximated by local \emph{renormalizable} operators
	that respect gauge invariance and, consequently, 
	can be absorbed into redefinition of Lagrangian parameters and fields. 

	In Section~\ref{sec:decoupling_escalar},  
	we discuss how relations between
	\MSbar-parameters and their counter-parts in \DR-scheme
	can be obtained on the same footing. 	
	The distinction between
	\MSbar\ and \DR\ essentially comes from the presence of
	so-called $\varepsilon$-scalars.  
	Since the scalars are unphysical, we may assume that they
	are heavy and decouple them in almost the same way as one decouples
	physical degrees of freedom. 
	As an example of the formalism, in Section~\ref{sec:qcd-qcd}, 	
	we consider two-loop matching of \DR\ QCD with \MSbar\ QCD.
	We show how known relations between \DR\ and \MSbar\ QCD parameters 
	(see, e.g., Ref.~\refcite{Harlander:2006xq}) are reproduced.		

	In Section~\ref{sec:qcd-sqcd}, we use 
	the described technique to calculate  
	a two-loop relation between 
	\MSbar\ and \DR\ running masses of the b-quark in QCD and
	SUSY QCD part of the MSSM. 
	In our approach, we decouple all heavy particles simultaneously (``common scale approach'' of
	Ref.~\refcite{Baer:2005pv}).
	For the TeV-scale SUSY it seems phenomenologically acceptable
	since the electroweak scale is usually used for matching.
	Simultaneous decoupling results 
	in a lengthy 
	final expression.
	As a consequence, only numerical impact of the result is presented.

	In our calculations we made use of FeynArts\cite{Hahn:2000kx} to 
	generate needed Feynman amplitudes. Since $\DR\to\MSbar$ transition
	requires explicit treatment of $\varepsilon$-scalars, 
	the interaction Lagrangian for the unphysical fields was
	implemented. Some details of the implementation can be found
	in a series of appendices.

\section{Decoupling of heavy particles and Large Mass Expansion}\label{sec:decoupling_framework}

In QCD and its supersymmetric extension it is
convenient to use mass-independent (minimal subtractions or \MS) 
renormalization schemes.
	In these schemes beta-functions and anomalous dimensions 
	have a very simple structure. 
	However, physical quantities expressed in terms of 
	the running parameters exhibit a nonanalytic logarithmic
	dependence on all mass scales of the theory.  
	
	If there is a big hierarchy between mass scales,
	it is not satisfactory, 
	since due to this nonanalytic mass dependence,
	a contribution of heavy degrees of freedom 
	to low-energy observables is not suppressed by the inverse power
	of the corresponding heavy mass scale.
	It is said that in \MS-schemes in contrast 
	to momentum subtraction schemes (MOM), the Appelquist-Carrazone 
	decoupling theorem\cite{Appelquist:1974tg} 
	does not hold.

	A proper way to overcome the above-mentioned difficulties of \MS-schemes
	is to use effective (low-energy) theories to describe physics 
	at relevant energy scales $E$.
	If at given energies heavy particles with mass $M > E$
	can only appear in virtual states
	one may use an effective field theory with
	the corresponding heavy fields  omitted.

	In a general case low-energy theories are not renormalizable. 
	Moreover, to reproduce physics close to the threshold 
	$E\lesssim M$ correctly, they should contain 
	infinitely many nonrenormalizable interactions
	parametrized by dimensionful couplings. 
	However, given a more fundamental theory 
	one can 
	relate all the couplings in the effective 
	low-energy Lagrangian 
	to fundamental parameters of the high-energy theory.
	Roughly speaking, one should calculate observables (or, more
	strictly, Green functions)
	in both theories and tune the parameters and field normalization 
	in the effective Lagrangian in 
	such a way that both results coincide in the region below 
	the threshold.

	This procedure is called matching. As a result
	of matching one expresses effective theory parameters as 
	functions of fundamental ones. 
	In practice, one cannot deal with an infinite number of interactions.
	So one usually performs \emph{asymptotic} expansion of Green 
	functions defined in high-energy theory
	in $E/M$ and demands that effective theory should correctly
	reproduce a finite order of the expansion considered.

	In some cases effective theories are used for energies
	far below the threshold so a contribution from 
	nonrenormalizable operators
	is usually suppressed by powers of $E/M$. One may even consider
	\emph{renormalizable} effective theory. In this case,
	the structure of low-energy Lagrangian differs from  the
	one of the fundamental theory only by omission of all the heavy
	fields and their interactions.
	
	One usually says that decoupling of heavy particles 
	is manifest if one can directly use the parameters 
	defined in a fundamental theory to calculate quantities
	within the effective theory (and vice versa). 
	In a momentum subtraction scheme 
	decoupling is obvious, since all the parameters
	are defined through Green functions evaluated at certain
	external momenta (less than $M$). 
	Since we demand that Green functions in 
	both theories coincide, all the parameters 
	defined in such a way also coincide.

	On the contrary, \MS-parameters are not related
	to Green functions evaluated at some finite momenta. 
	Thus, one does not expect that \MS-parameters have the same value
	in both the theories.
	In this sense decoupling in \MS-schemes does not hold. One should
	manually tune the parameters defined in such a scheme.
	
\begin{figure}[h!]
	\begin{center}
		\includegraphics{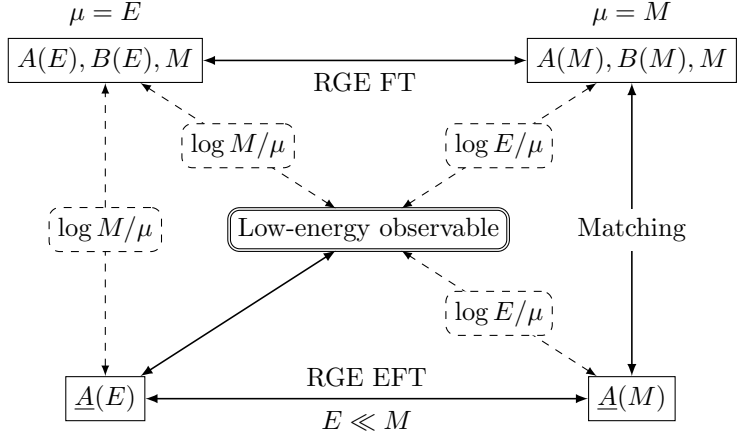}
	\end{center}
	\caption{The diagram shows various relations between full theory (FT) running parameters ($A,B,M$),  
	the parameters $\underline{A}$ of the effective theory (EFT), and low-energy observables.
	Renormalization group equations (RGE) together with matching at the scale $M$ allows one
	to avoid the appearance of large logarithms (see dashed boxes) in calculation.
	Solid lines correspond to relations that do not introduce large logarithms explicitly.}
	\label{fig:eft_vs_ft}
\end{figure}

	How does the approach based on effective theories 
	help one to avoid the appearance of large logarithms 
	for $E \ll M$ in a calculation?
	The trick is to separate  
	$\log M/E$ into $\log M/\scale$ and $\log E/\scale$, 
	where $\scale$ is an arbitrary separation scale which in \MSbar-scheme is
	naturally equal to the 'tHooft unit mass.
	Then one may absorb $\log M/\scale$ into low-energy parameters
	of the effective theory by a matching procedure.
	A typical relation between the parameters is
\begin{equation}
	\underline{A}(\scale) = A(\scale) \times \zeta_A (A(\scale), B(\scale), M, \scale),
	\label{decouling_constants:definition}
\end{equation}
	where $\underline{A}$ is a dimensionless parameter of the effective theory\footnote{In what follows
	we underline effective theory parameters and fields} , $A$ denotes
	its counter-part in the high-energy theory, and $B$ are other dimensionless fundamental theory 
	parameters.
	The function $\zeta_A$ is called decoupling 
	constant for $A$. 
	Generalization of \eqref{decouling_constants:definition} to dimensionful parameters is straightforward. 
	However, one should keep in mind that the crucial property of  $\zeta_A$
	is its independence of  small energy scales. 
	One calculates $\zeta_A$ in perturbation theory and  
	at the tree level $\zeta_A = 1$. 
	Given $\scale\sim M$ one avoids 
	large logarithms in \eqref{decouling_constants:definition}.
	If one knows fundamental parameters at the scale $\scale \sim M$, one can find the values of the effective-theory
	parameters at the same scale. Clearly, direct application of the parameters $\underline{A}  (\scale \sim M)$ in low-energy
	theory again introduces large logarithms $\log E/\scale \sim \log E/M$. 
	However, in the effective theory it is possible to sum these logarithms by the renormalization group method, 
	i.e., going from $\underline{A}(M)$ to $\underline{A}(E)$ (see Fig.~\ref{fig:eft_vs_ft}) .

	Actually, one usually reverses the reasoning. Typically, fundamental theory parameters are unknown (especially, the value of $M$), 
	but one knows the value of $\underline A(E)$ normalized at some low-energy scale $E$. 
	Direct application of \eqref{decouling_constants:definition}
	again introduces large logarithms in the right-hand side,
	$\log M/\scale \sim \log M/E$. As in the previous case, one should use renormalization group equations
	defined in the effective theory to evaluate $\underline A(\scale)$  at $\scale\sim M$.
	Consequently, relation \eqref{decouling_constants:definition} 
	can be interpreted as a constraint on the fundamental theory parameters.  
	Usually, largest variations of the right-hand side of \eqref{decouling_constants:definition} come from
	variations of $A$ parameter, so one says that the value of $A(\scale)$ is extracted from $\underline A(\scale)$.
	It is this type of relations we are interested in.
	Let us describe the procedure that one can use to calculate
	decoupling (matching) corrections. 

	Consider the Lagrangian of a full theory, $\L_{full}$. For the moment, we do not
	specify it explicitly. The crucial property of the theory is that it describes not
	only gluons and quarks (light fields denoted by $\phi$), but also 
	heavy fields $\Phi$ with typical masses $M$: 
\begin{equation} \label{lag:full_general}
	\L_{full}  =  \L_{QCD}\Bigl( \phi,\, g_s,\, m,\,\xi\Bigr) 
		+ \Delta \L \Bigl( \phi,\, \Phi,\, g_s,\, m,\, M,\dots).
\end{equation}
	Here $\phi \equiv \left\{G_{\scale},\, q_{L,R},\, c\right\}$ are gluon, quark, and ghost fields,
	respectively. The strong gauge coupling is denoted by $g_s$, $m$ 
	corresponds to quark masses and $\xi$ is a gauge-fixing parameter.
	It should be noted that $\Delta \L$ represents the Lagrangian for heavy fields
	and contains kinetic terms for $\Phi$ together with various interactions\footnote{In this paper, we only
	consider strong interactions between all fields of the fundamental theory}. 	
	The QCD Lagrangian $\L_{QCD}$ has the usual form:
\begin{eqnarray} \label{lag:QCD}
	\L_{QCD} & = & - \frac{1}{4} F^a_{\mu\nu} F_a^{\mu\nu}  +  \overline{q} \left(i \hat D -  m \right) q, \nonumber\\
	 & - &  \frac{1}{2 \xi} \left( \partial_\mu G_a^\mu \right)^2
	- \overline{c}^a \partial^{\mu} \left( \partial_\mu \delta^{ab} + g_s f^{abc} G^c_{\mu} \right) c^b,
\end{eqnarray}
	where
\begin{eqnarray*}
	F^a_{\mu\nu} & = & \partial_\mu G^a_{\nu} - \partial_\nu G^a_{\nu} - g_s f^{abc} G^b_\mu G^c_\nu,  \\
	D_\mu & = & \partial_\mu + i g_s G_\mu,\quad G_{\mu} = T^a G^a_{\mu}, \\
	\left[T^a,T^b\right] & =  & i f^{abc} T^c, 
	\quad \Tr \, T^a T^b = T_F \delta^{ab}, \quad T_F = 1/2
\end{eqnarray*}
	and for simplicity we omit summation over quark flavours. 
	For energies below the threshold $E < M$ one is interested in Green functions
	with light external fields
\begin{eqnarray} 
	\langle T \phi(x_1)\dots \phi(x_n) \rangle^{\L_{full}} 
& \equiv  &  
	 \int \mathcal{D} \phi \, 
	\Bigl( \, \phi(x_1)\dots\phi(x_n) \Bigr) 
	\exp i \int d x \,
	\L_{full}\left(\phi, \Phi\right) \\
	G(q_1, \dots q_n, m, M) 
& \equiv & 
	\int \Bigl[ \prod\limits_{i=1}^{n} d x_i \Bigl]
	\exp \, i \Bigl[ \sum\limits_{i=1}^{n} q_i x_i \Bigr]
	\times \langle T \phi(x_1)\dots \phi(x_n) \rangle^{\L_{full}}, 
\label{green_function:full}
\end{eqnarray}
	renormalized in a minimal scheme.
	Let us consider the expansion of \eqref{green_function:full} in the inverse powers of $M$
	(Large Mass Expansion or \LME).
	The leading order of the expansion 
	can be written in the following form:
\begin{eqnarray} \label{matching:gen_condition}
	\langle T \phi(x_1)\dots \phi(x_n) \rangle^{\L_{full}(\phi,\Phi)} 
	& \underset{M\to \infty}{\simeq}& 
\langle T \phi(x_1)\dots \phi(x_n) \rangle^{\L_{eff}}(\phi)  
	+ \mathcal{O}(M^{-1})
\end{eqnarray}
	or in momentum space
\begin{eqnarray}\label{matching:gen_condition_momentumspace}
	G(q,m,M) 
& \underset{M\to \infty}{\simeq}& 
	\underline{G}(q,m,M) + \mathcal{O}(M^{-1}), \nonumber\\ 
	\underline{G}(q_1, \dots q_n, m, M) 
& \equiv & 
	\int \Bigl[ \prod\limits_{i=1}^{n} d x_i \Bigl]
	\exp \, i \Bigl[ \sum\limits_{i=1}^{n} q_i x_i \Bigr]
	\times \langle T \phi(x_1)\dots \phi(x_n) \rangle^{\L_{eff}}
	.
\end{eqnarray}
	Here
\begin{align}
	\L_{eff}  = & \L_{QCD}(\phi)+ \delta\L_{QCD}(\phi) \nonumber\\
\delta \L_{QCD}
 = & 
	-\frac{1}{2} \delta \zeta_{G} \, 
		\left( \partial_\mu G^a_\nu - \partial_\nu G^a_\mu \right) 
		\partial_\mu G^a_\nu 
 +  
	\delta \zeta_{3G} \left( g_s f^{abc} (\partial_{\mu} G^a_{\mu}) G^b_{\mu} G^c_{\mu} \right) \nonumber\\ 
& -   
	\frac{1}{4} \delta \zeta_{4G} \left( g_s^2 f^{abe} f^{cde} 
	G^a_{\mu} G^b_{\nu} G^{c}_{\mu}  G^d_{\mu}  \right) 
 +  
	\delta \zeta_{c} \left(\partial_\mu \overline{c}^a \partial^\mu c^a \right) 
 \nonumber\\
& +   
	  \delta \zeta_{q_L} \left( 
	\overline{q}_L \, i \hat \partial \,q_L \right)
	+  \delta \zeta_{q_R} \left(
	\overline{q}_R \, i \hat \partial \,q_R \right)   
 	-  
	\delta \zeta_{s} \left( m \, \overline{q} q  \right)
	\nonumber\\
& +  
	\delta \zeta_{cGc}\left[ g_s f^{abc} 
	\left(\partial^{\mu} \overline{c}^a\right) c^b G^c_{\mu} \right] 
  -  \sum\limits_{l=L,R}\delta \zeta_{q_lGq_l}\left( g_s
	\, \overline{q}_l \, T^a \hat G^a \, q_l \right) 
	\label{matching:QCD_eff_2loop}
\end{align}
	and coefficients $\delta \zeta_i\equiv \zeta_i - 1$ in \eqref{matching:QCD_eff_2loop} are
	functions of $M$ with logarithmic leading behaviour as $M\to\infty$. 
	The form of asymptotic expansion 
	\eqref{matching:gen_condition_momentumspace} 
	represents the perfect factorization property
	\cite{Tkachov:1991ev}, since heavy ($M$) and light parameters 
	($m$) are fully factorized.  We consider only 
	leading term in the expansion\footnote{One may increase the accuracy of
	the expansion in \eqref{matching:gen_condition} by adding to $\delta
	\L$ nonrenormalizable local operators built of $\phi$ with the
	coefficients ${\mathcal O}(M^{-a}),~a>1$}, so 
	all operators in $\delta \L_{QCD}$ are
	renormalizable. 
	Therefore, one can rescale
	the light fields
\begin{equation} \label{decoupling:field_general}
	\myunderline{\phi}  =  \zeta_{\phi}^{1/2} \phi, \qquad
	\zeta_{\phi}  =  1 + \delta \zeta_{\phi}
\end{equation}
	and write the effective theory Lagrangian in terms of $\myunderline{\phi}$
\begin{eqnarray}\label{effective_qcd}
	\L_{eff}(\phi,\, g_s,\, m,\, \xi) & = & 
	\L_{QCD}\Bigl(
	\myunderline \phi,\,
		\myunderline{g_s},
		\myunderline{m},
		\myunderline \xi\Bigr), 
\end{eqnarray}
	where we also introduce  new parameters 
	which are related to the initial ones \eqref{lag:full_general} 
	by means of decoupling constants ($l=L,R$) 
\begin{subequations}
\label{decoupling:general}
\begin{equation*}
\myunderline{g_s}  =  \zeta_{g_s} g_s, \qquad 
\myunderline{m}  =  \zeta_{m} m,  \qquad
\myunderline{\xi}  =  \zeta_{G} \, \xi,
\end{equation*}
	where
\begin{align}
	\zeta_{g_s}  =	 \zeta_{3G} \zeta_{G}^{-3/2} 
=	\zeta_{4G}^{1/2} \, \zeta_{G}^{-1}
= & \, 	\zeta_{cGc} \, \zeta_{c}^{-1} \, \zeta_{G}^{-1/2}
= 	\zeta_{q_lGq_l} \, \zeta_{q_l}^{-1} \, \zeta_{G}^{-1/2}
	\label{decoupling:gs_general}, \\
	\qquad \zeta_{m} = & \, \zeta_s \, \zeta_{q_L}^{-1/2} \, \zeta_{q_R}^{-1/2}. 
	\label{decoupling:mass_general}
\end{align}
\end{subequations}
	Due to the gauge invariance one should obtain the same result for $\zeta_{g_s}$ 
	in \eqref{decoupling:gs_general} 
	calculated from different vertices. 
	Moreover, since dimensional regularization does not violate 
	the gauge invariance, the longitudinal part of the gluon propagator 
	is not renormalized. As a consequence, for the gauge-fixing
	parameter $\xi$ 
	one introduces the same decoupling constant as for the gluon
	field.
	According to \eqref{effective_qcd}, one can identify underlined 
	parameters with those of QCD. Heavy degrees of freedom
	are said to be ``decoupled''. 

	We should stress 
	that $g_s$, $m$, and $\xi$ in the previous formulae are 
	renormalized parameters and all the decoupling constants 
	are finite. Evaluation of the constants for $g_s$ and $m$ 
	requires a comparison 
	of certain Green functions calculated
	with $\L_{eff}$ with the lowest order expansion of the 
	same functions calculated with $\L_{full}$. 
	The matching is performed order by order in perturbation theory.
	Introducing 
\begin{equation} \label{decoupling:loop_expansion}
	\zeta = 1 + \sum\limits_{l=1}^{\infty} \delta \zeta^{(i)}
\end{equation}
	one can write the $L$-loop contribution to the decoupling constant
	$\zeta_v$ for each vertex $v$ from \eqref{matching:QCD_eff_2loop}:
\begin{eqnarray} \label{matching:main_formula_practical}
	\dzeta{v}{L} (M) & = & {\mathcal{P}}_v \circ
	\Bigl[
	\mathrm{As} \circ \Gamma^{(L)}_v(q,m,M) 
	- \myunderline{\Gamma_v}^{(L)}(q,m,M) 
	\Bigr].
\end{eqnarray}
	Here $\Gamma^{(L)}_v$ denotes the $L$-loop 
	contribution to the renormalized  one-particle-irreducible
	(1PI) Green function 
	that corresponds to the vertex $v$.
	The operator $\mathrm{As}$ performs
	asymptotic expansion\cite{Tkachov:1991ev} of $\Gamma^{(L)}_v$ calculated with $\L_{full}$
	up to the leading term in the inverse mass $M$. 
	For calculation of $\myunderline{\Gamma_v}^{(L)}$
	one uses the $(L-1)$-loop effective theory Lagrangian 
	that differs from \eqref{matching:QCD_eff_2loop} 
	only by omission of all the terms 
	in \eqref{decoupling:loop_expansion} with $i\geq L$.
	The appropriate projector ${\mathcal P}_v$ applied to  
	the Green function extracts the needed coefficient 
	in front of the considered tensor (Lorentz, color, etc) structures
	(see examples below).
	All nonanalytical dependence on low mass scales is canceled 
	in the left-hand side of 
	\eqref{matching:main_formula_practical}	
	leading to 
\begin{eqnarray} \label{matching:main_formula_practical_taylor}
	\dzeta{v}{L} (M) & = & {\mathcal{P}}_v \circ
	\mathcal{T} \circ
	\Bigl[
	\Gamma^{(L)}_v(q,m,M) 
	- \myunderline{\Gamma_v}^{(L)}(q,m,M) 
	\Bigr],
\end{eqnarray}
	where $\mathcal T$ performs Taylor expansion in small
	mass $m$ and external momenta $q$. 
		
	The procedure described above is straightforward, 
	since it deals with the well-defined finite quantities 
	but not the optimal one. 
	Formulae \eqref{matching:main_formula_practical} and
	\eqref{matching:main_formula_practical_taylor} require
	evaluation of the Green functions within both the theories.

	Let us recall that asymptotic expansion 
	of a Feynman integral constists of the naive part
,	and the subgraph part. The naive part corresponds to 
	Taylor expansion of the integrand in small parameters 
	that cannot give rise to a nonanalytical
	dependence on low mass scales. 
	Subgraphs restore missing terms in the result.
	The calculation of asymptotic
	expansion can be rearranged in such a way (see below) 
	that subgraphs
	of various diagrams contributing to the first term 
	of \eqref{matching:main_formula_practical}
	cancel the $M$-dependent contribution 
	to the second term in the squared brackets.
	Taking into account that diagrams with all vertices  
	coming from the QCD part of the Lagrangian contribute identically
	to both the terms of \eqref{matching:main_formula_practical}, 
	decoupling constant calculations 
	can be reduced to the evaluation of the naive part of
	\LME\ of the diagrams with at least one heavy
	line. 

	There is another issue that has to be pointed out. 
	Taylor expansion of the integrand may produce spurious 
	IR divergences which can be avoided by a proper redefinition
	of dangerous terms 
	in the sense of distributions\cite{Tkachov:1997gz}.
	In a dimensionally regularized form of the expansion 
	the spurious divergences are canceled by the UV-divergent terms
	coming from the subgraphs. The rearrangement mentioned above
	is nothing else but addition of a necessary counter-term to
	the naive expansion and subtraction of the same expression 
	from the subgraphs\cite{Tkachov:1991ev}. 

	A nice trick (see, e.g., Ref.~\refcite{Chetyrkin:1997un}) can be used
	to maintain the rearrangement automatically. One introduces
	decoupling constants for bare parameters $A_0$  
\begin{equation} \label{decoupling:bare}
	\myunderline{A_0}  =  \zeta_{A,0} \,  A_0  
\end{equation}
	and carries out matching at
	the bare level. In this case, naive Taylor expansion 
	in small parameters is used to calculate the $L$-loop contribution
	to the bare decoupling constant 
\begin{eqnarray} \label{matching:main_formula_practical_bare}
	\dzeta{v,0}{L} (M) & = & {\mathcal{P}}_v \circ
	\mathcal{T} \circ \Gamma^{(L)}_{v,0}(q,m,M).
\end{eqnarray}
	Then by the same formulae \eqref{decoupling:general} one obtains
	relations \eqref{decoupling:bare} between bare parameters
	of the low- and high-energy theories.
	Definitely, this calculation introduces spurious IR divergences.
	However, they are completely canceled when one renormalizes the
	left- and right-hand sides of \eqref{decoupling:bare} in
	\MS-scheme
\begin{eqnarray}
\myunderline{A_0} & = & \myunderline{Z_A} \, (\myunderline{A}) \,
	\myunderline{A}, \qquad
A_0  =  Z_A \, (A, B) \, A, \\
	\zeta_{A}
& = & 
\Bigl[ Z_A(A,\, B) \Bigr]
\Bigl[\,\myunderline{Z_A}\,(\myunderline{A}) \,\Bigr]^{-1}
	\,
	\zeta_{A,0}(Z_A \, A, Z_B \, B, \, Z_M M)
	. \label{decoupling:renormalized}
\end{eqnarray}
	Here we emphasized that renormalization constants 
	$Z_A$ and $\myunderline{Z_A}$ are defined in different 
	theories and depend on the parameters of the full ($A,B,M$)
	and effective theories ($\myunderline{A}$), respectively.
	Since $\underline{A}$ enters into the right-hand side of 
	\eqref{decoupling:renormalized}, this is an equation
	that should be solved in perturbation theory.

	Finally, we want to make a remark that a decoupling relation 
	between \MS-parameters can also be found\cite{Bernreuther:1981sg} 
	by considering  momentum space (''physical'') subtractions 
	as an intermediate step. Indeed, the parameter $A_{\MOM}(Q)$
	defined in \MOM-scheme at some scale $Q \ll M$ can be expressed
	either in terms of $\underline{A}$ (effective theory) 
	or in terms of $A,B$ and $M$ (full theory)
\begin{eqnarray} \label{decoupling:physical}
	A_{\MOM}(Q) & = & \myunderline{f} (\myunderline{A}, Q) 
		      = f(A, B, M, Q). 
\end{eqnarray}
	As it should be, it turns out 
	that the relation between $\underline{A}$ and 
	the parameters of the full theory 
	do not depend on $Q$.

\section{Transition from \DR\ to \MSbar\ by decoupling of $\varepsilon$-scalars}\label{sec:decoupling_escalar}

In dimensional regularization (DREG), the number of space-time dimensions is altered from four to
$d=4-2\varepsilon$ which renders the loop integrations finite.  It is
clear, however, that if DREG is applied to a four-dimensional
supersymmetric theory, the number of bosonic and fermionic degrees of
freedom in supermultiplets is no longer equal, such that supersymmetry
is explicitly broken.  In order to avoid this problem,
Dimensional Reduction  has been suggested as an alternative
regularization method\cite{Siegel:1979wq}. Space-time is compactified to
$d=4-2\varepsilon$ dimensions in DRED, such that the number of vector
field components remains equal to four. Momentum integrations are
$d$-dimensional, however, and divergences are parametrized in terms of
$1/\varepsilon$ poles, just like in DREG. Since it is assumed that
$\varepsilon> 0$, the four-dimensional vector fields can be decomposed in
terms of $d$-dimensional ones plus the so-called $\varepsilon$-scalars.  The
occurrence of these $\varepsilon$-scalars is, therefore, the only
difference between DREG and DRED, so that all the calculational
techniques developed for DREG are applicable also in DRED\cite{Harlander:2006xq}.
	
	Dimensional reduction of the four-dimensional $\L_{QCD}$ leads to the following
	regularized form of QCD Lagrangian (see \ref{sec:escalar-qcd})
\begin{align}
	\L_{QCD}  \to  &\, \L^{4-2\varepsilon}_{QCD} + \delta \L^{\varepsilon}_{QCD} \nonumber\\
	\delta \L^{\varepsilon}_{QCD}  =   
		 & - \frac{1}{2} 
		 \left(D_\mu W^i\right)_a  
		 \left(D^\mu W_i\right)_a  
		 \nonumber\\ 
	& - \frac{1}{4} 
		    g_s^2 f^{abc} f^{a\tilde b \tilde c}    
		W^b_i W^c_j W^{\tilde b}_i W^{\tilde c}_j 
	- g_s \overline q \gamma^i W^a_i T^a q \label{lag:qcd-escalar},
\end{align}
	where $W^a_i$ corresponds 
	to $\varepsilon$-scalar fields and the indices $i,j$ belong
	to the (space-like) $2\varepsilon$-subspace 
	of the four-dimensional world. 
	In nonsupersymmetric models there are several issues 
	related to this Lagrangian. 
	First of all, 
	the last two terms in \eqref{lag:qcd-escalar} 
	are gauge-invariant separately\cite{Jack:1993ws},
	and there is no symmetry that guarantees the same renormalization of the couplings 
	that parametrize these vertices (see \ref{sec:escalar-qcd}). 
	This leads to complications in the renormalization group analysis,
	since the running of the couplings \eqref{lag:qcd-escalar}
	is different.
	Secondly, 
	since $W_i$ are scalars, 
	all massive particles that couple to them contribute to 
	the unphysical $\varepsilon$-scalar mass $\meps^2$.
	In order to solve the first mentioned problem, one must 
	introduce so-called evanescent\cite{Jack:1993ws} 
	couplings for each vertex. 
	To
	the second problem there are two approaches.
	One may either introduce $\varepsilon$-scalar mass explicitly in 
	the Lagrangian \eqref{lag:qcd-escalar} and renormalize 
	it minimally (\DR-scheme) or use nonminimal counter-term to 
	subtract radiative corrections to $\meps^2$ at each order of perturbation theory\cite{Avdeev:1997sz}. 
	It turns out that these prescriptions 
	give rise to the same final answer for 
	the QCD observables. 
	We have checked this explicitly by considering two-loop pole mass\cite{Avdeev:1997sz,Marquard:2007uj} 
	of the quark in the \DR\ QCD.	

	In the context of SUSY QCD part of the MSSM the situation is different. 
	All the dimensionless evanescent couplings are related to the gluon
	couplings by SUSY and, therefore, are not independent. This circumvents the first problem.
	However, two mentioned renormalization conditions 
	for $\meps^2$ produce different answers.  
	For example, in \DR\ the pole mass of a scalar quark superpartner (squark) exhibits a
	quadratic dependence on $\meps^2$. 
	The authors of Ref.~\refcite{Jack:1994rk} proposed to redefine 
	running squark masses $m^2_{\tilde q}$ to absorb the unphysical contribution  
	($\DR'$-scheme). At the one-loop level\footnote{Two-loop result can be found in Ref.~\refcite{Martin:2001vx}}	
	one has
\begin{equation} \label{MSDR:first_step}
	\left( m^2_{\tilde q} \right)_{\DR'} =  \left( m^2_{\tilde q} \right)_{\DR}  - 2 C_F \frac{\as}{4\pi} \, \meps^2.
\end{equation}
	After such a redefinition one obtains the result which 
	is independent of $\meps^2$.	
	The new scheme is equivalent to the prescription 
	with a nonminimal counter-term\cite{Avdeev:1997sz}. This
	statement was also checked explicitly by considering heavy quark pole mass\cite{Bednyakov:2002sf} 
	in SUSY QCD as an observable. 

	Formula \eqref{MSDR:first_step} and the reasoning 
	that was used to obtain it allows one to interpret \eqref{MSDR:first_step} 
	as a first step towards decoupling of $\varepsilon$-scalars 
	in the sense described in the previous section.  
	Leading (but unphysical) $(\meps^2\to\infty)$ corrections to the pole mass of the squark 
	are absorbed into redefinition of the corresponding mass parameter. 
	One may go further and decouple $\varepsilon$-scalars 
	completely. It seems useless in the context of SUSY QCD,
	since without $\varepsilon$-scalars one loses the
	advantages of DRED. 
	However, it makes sense in the problem described in the paper.
	For nonsupersymmetric models \MSbar-scheme is natural
	in the sense that contrary to \DR\, it does not require 
	introduction of evanescent couplings.
	Given the procedure (see Sec.~\ref{sec:decoupling_framework}),
	not only physical degrees of freedom can be decoupled but also 
	unphysical $\varepsilon$-scalars. 
	This leads to a direct relation between $\MSbar$-parameters
	of the effective theory and $\DR$-parameters of the full theory.

	Using these simple arguments we calculate the relation
	between the $b$-quark running mass defined in the \MSbar\ QCD 
	and its counter-part 
	in $\DR$ SUSY QCD. 
	Before going to the final result, 
	in the next section we want to demonstrate how
	known relations between \DR\ and \MSbar\ QCD parameters are
	reproduced.

\section{Toy example: matching \MSbar\ QCD with \DR\ QCD}\label{sec:qcd-qcd}
	
	Let us consider a model with $(n_f-1)$ massless quarks 
	and only one massive quark with mass denoted by  $m$. 
	Thus, the task is to find two-loop relations of the following type:
\begin{equation} \label{toy:definitions}
	g_s^{\MSbar}   =  g_s^{\DR} \times 
		\zeta_{g_s} (\as^{\DR}, \ay^{\DR}), \qquad
	m^{\MSbar}   =  
	m^{\DR} \times \zeta_{m} (\as^{\DR}, \ay^{\DR})
\end{equation}
	where 
\begin{equation*}
	\as = \frac{g_s^2}{4 \pi}, 
	\ay = \frac{g_y^2}{4 \pi}. 
\end{equation*}
	In \eqref{toy:definitions} superscript tells us what kind of 
	renormalization scheme is used and $g_y$ is 
	the evanescent coupling for 
	$\varepsilon$-scalar interaction 
	with quarks (see~\eqref{feynman:escalar-quarks}).
	Usually, \eqref{toy:definitions} are solved in perturbation
	theory to obtain $m^{\DR}$ and $g_s^{\DR}$ as functions
	of $\MSbar$-parameters and evanescent ones. 
	However, we use the form \eqref{toy:definitions}, since
	it is directly related to matching.

	In this section, we consider ``high-energy'' theory with a
	Lagrangian $\Delta \L$ \eqref{lag:full_general}
\begin{align}
	\Delta \L  = &   
		 - \frac{1}{2} 
		 \left(D_\mu W^i\right)_a  
		 \left(D^\mu W_i\right)_a  
		 +
		 \frac{1}{2} \meps^2 W^i_a W_i^a \nonumber\\
		 & - 
	\frac{1}{4} \sum\limits_{r=1}^3 \lambda_r H^{abcd}_r 
	W_i^a W_j^c W_i^b W_j^d
	- g_y \overline q \gamma^i W^a_i T^a q 
	\label{full_lag:qcd-escalar}
\end{align}
	that describes ``heavy'' degrees of freedom. 
	In \eqref{full_lag:qcd-escalar} the mass
	of the unphysical scalars $\meps^2$ is explicitly introduced
	together with evanescent couplings $g_y$ and $\lambda_r$. 
	Tensors $H_r^{abcd}$ have certain symmetric properties 
	(see~\ref{sec:escalar-qcd}) and
	define a color structure of the four-vertex. 
	
	First of all, to calculate bare decoupling constants 
	that correspond to \eqref{matching:QCD_eff_2loop}
	we consider bare 1PI Green functions and their Taylor expansion
	in small masses and external 
	momenta \eqref{matching:main_formula_practical_bare}.
	In the \DR\ QCD left- and right-handed quarks are renormalized 
	in the same way, so let us introduce $\delta \zeta_{q} = \delta \zeta_{q_L} = \delta \zeta_{q_R}$
	and $\delta \zeta_{qGq} = \delta \zeta_{q_L G q_L} = \delta \zeta_{q_R G q_R}$.

	For calculation of $\delta \zeta_{q,0}$ and $\delta \zeta_{s,0}$ 
	quark self-energy $\Gamma_{q}$ is used
\begin{eqnarray}
i \Gamma_{q} (\hat p,m,\meps^2) & = & 
	\Sigma_v(p^2,m^2,\meps^2) \, \hat p 
	+ \Sigma_s(p^2,m^2,\meps^2) \, m \nonumber\\
\delta \zeta_{q} & = & 
	\left. \left(-i  \frac{1}{4 \, n_c \, p^2} \times 
	\Tr  \, \hat p \, \Gamma_{q} \, \right)\right|_{p,m=0}
\!\!\!\!\!\!
	= - \Sigma_v(0,0,\meps^2) 
\label{bare_dec:qv}\\
\delta \zeta_{s} & = & 
	\left. \left(+i \frac{1}{4  \, n_c \, m }  \times
	\Tr \; \Gamma_{q} \;  \right)\right|_{p,m=0}
\!\!\!\!\!\!
	= \Sigma_s(0,0,\meps^2)
\label{bare_dec:qs},
\end{eqnarray}
	where the trace is taken both over spinor and color 
	indices and $4 n_c$ appears in the denominator
	because of chosen normalization 
	($\Tr\,\mathbf{1} = 4$ for Dirac algebra and  $\Tr\,\mathbf{1} = n_c$ 
	for color algebra).
	
	Due to the gauge invariance not all the parameters
	from \eqref{matching:QCD_eff_2loop} are needed to 
	find $\delta \zeta_{g_s,0}$. The simplest choice is to use
	the ghost-gluon vertex $\Gamma_{cGc}$
\begin{eqnarray} \label{bare_dec:cGc}
g_s \, \delta \zeta_{cGc} & = &  \left. \left(
	\frac{1}{n_g} f^{abc} \times \frac{k_{\mu}}{k^2} \times  
	\Gamma^{abc,\, \mu}_{cGc} \right) \right|_{p,m=0},
\end{eqnarray}
	where $a,b,c$ are gluon indices, $f^{abc}$ corresponds
	to $SU(3)$ structure constants, $n_g=8$ is the number
	of gluons, $p$ denotes all external momenta 
	and $k_{\mu}$ is the momentum 
	of incoming antighost (cf. \eqref{matching:QCD_eff_2loop}).
	One also needs to consider gluon and ghost self-energies
\begin{eqnarray}
i \Gamma^{\mu\nu,\,ab}_{G}(p,m^2,\meps^2) & = & 
- \delta^{ab} \, \left(g^{\mu\nu} p^2 - p^{\mu} p^{\nu} \right) \,
\Pi_G(p^2,m^2,\meps^2), 
\label{self-energy:gluon}\\
i \Gamma^{ab}_{c}(p^2,m^2,\meps^2) & = & 
\delta^{ab} \, p^2 \, \Sigma_{c}(p^2,m^2,\meps^2) 
\label{self-energy:ghost}
\end{eqnarray}
	so 
\begin{eqnarray}
\delta \zeta_{G} & = & 
\left. \left(
		+ i \frac{\delta^{ab}}{n_g} \times 
		\frac{1}{d-1}\left( g_{\mu\nu} - \frac{p_\mu p_\nu}{p^2} \right)
		\frac{1}{p^2} \times \Gamma^{\mu\nu,\,ab}_{G} \right) 
		\right|_{p,m=0} \!\!\!\!\!\!
	= - \Pi_{G}(0,0,\meps^2)
	\label{bare_dec:G}, \\
\delta \zeta_c & = & 
\left. \left(
		-i \frac{\delta^{ab}}{n_g} \times 
		\frac{1}{p^2} \times \Gamma^{ab}_{c} \right) 
		\right|_{p,m=0} \!\!\!\!\!\! 
	= - \Sigma_c(0,0,\meps^2)
\label{bare_dec:c},
\end{eqnarray}
	where $d=4-2\varepsilon$, since we use $d$-dimensional metric tensor
	in \eqref{bare_dec:G}. To check the final result for
	$\delta \zeta_{g_s,0}$, we also use the gluon-quark vertex $\Gamma_{qGq}$
\begin{equation} \label{bare_dec:qGq}
g_s \delta \zeta_{qGq}  =  \left. \left(
	+ i \frac{1}{4 n_c C_F} 
	\times \Tr \; \gamma^{\mu} \Gamma_{qGq}^{\mu,a} T^a  
	\right) \right|_{p,m=0},
\end{equation}
	where $C_F = 4/3$ is a casimir of $SU(3)$ and again the trace is taken over both spinor and color indices.

	Direct evaluation of the diagrams that contribute to the 
	one-loop decoupling constants gives
\begin{eqnarray}
	\dzeta{qGq}{1} & = & \ay C_F  
	\left( 1 + \varepsilon 
	\left( \frac{1}{2} - \Lmes \right)
	\right) \label{dec:qGq1}, \\
	\dzeta{q}{1} & = & \ay C_F   
	\left( 1 + \varepsilon 
	\left( \frac{1}{2} - \Lmes \right)
	\right) \label{dec:q1}, \\
	\dzeta{G}{1} & = & \frac{1}{3} \, \as C_A     
	\left( 1 - \varepsilon \Lmes \right) \label{dec:G1} \\
	\dzeta{s}{1} & = & 2 \ay C_F     
	\left( 1 + \varepsilon \left( 1 - \Lmes \right)\right) 
	\label{dec:s1}, \\
\dzeta{m}{1} & = & \dzeta{s}{1} - \dzeta{q}{1} = \ay C_F     
\left( 1 + \varepsilon \left( \frac{3}{2} - \Lmes \right)\right) 
	\label{dec:m1}, \\
\dzeta{g_s}{1} & = & \dzeta{qGq}{1} - \dzeta{q}{1} - \frac{1}{2} \dzeta{G}{1}  \nonumber\\
	       & = &  \dzeta{cGc}{1} - \dzeta{c}{1} - \frac{1}{2} \dzeta{G}{1} =  
- \frac{1}{6} \as C_A \left( 1 - \varepsilon \Lmes \right).
\end{eqnarray}
	Here all the coupling constants are considered to be defined in \DR, 
	$C_A=3$ is another casimir of $SU(3)$,  $\Lmes \equiv \log \meps^2/\scale^2$ and $\dzeta{cGc}{1} = \dzeta{c}{1} = 0$. 
	Notice
	that at the one-loop level bare decoupling constants for the mass and 
	the gauge coupling
	are finite as $\varepsilon \to 0$ and exhibit a dependence on $\Lmes$ only when $\varepsilon \neq 0$. 
	As usual, since we want to consider two-loop matching we keep terms that are linear in $\varepsilon$.

	Two-loop decoupling corrections look like
\begin{eqnarray}
\dzeta{qGq}{2} & = & 
	\left(\frac{1}{\varepsilon} - 2 \Lmes \right)
	\left(
	\ay C_F \left[ \vphantom{\frac{1}{2}}
	3 \as \,C_F - \ay \, \left(
	 2 C_F - C_A + n_f T_F \right)		
	 \right]
	+ \frac{1}{8} \as^2 \,C_A^2  
	\right) \nonumber \\
	& + & \ay C_F \left( 7 \lambda_3 
		+ 20 \lambda_2 
		- \lambda_1 C_A 
		+ \as \left(
		\frac{3}{2}  C_A 	
		+ C_F 
		\right)
		\right) \nonumber\\
	&  -  &  
	\frac{1}{4} \as^2 C_A  \left(
	\frac{11}{12} C_A + C_F \right)
	 -   
	\frac{1}{2} \ay^2 \, C_F 
	\left( C_F +  n_f T_F  
	\right)   
	\label{dec:qGq2} \\
\dzeta{c}{2} & = & \frac{1}{4} \as^2 C_A^2 
	\left(
	- \frac{1}{2 \varepsilon} 
	+ \frac{11}{24} 
	+ \Lmes 
	\right) \label{dec:gh2} \\
\dzeta{q}{2} & = & 
	\left(\frac{1}{\varepsilon} - 2 \Lmes \right) \ay C_F 
	\left[ \vphantom{\frac{1}{2}}
	3 \as \, C_F - 
	\ay \left( 2 C_F - C_A  + n_f T_F  \right)
	\right] \nonumber\\
& + &  \ay C_F \left( 
	7 \lambda_3  
	+ 20 \lambda_2 
	- \lambda_1  C_A 	
	+ \as \left( 
		\frac{3}{2}  C_A
		+ C_F
	\right)
	\right) \nonumber\\
& - & \frac{1}{2} \ay^2 C_F 
	\left(C_F + n_f T_F   
	\right) 
-\frac{1}{4} a_s^2 C_F C_A \label{dec:q2}  \\
\dzeta{s}{2} & = & 	
	\left( \frac{1}{\varepsilon} - 2 \Lmes \right) C_F  \left(
	2 \ay \, C_F \left[ \vphantom{\frac{1}{2}}
	3 \as C_F - \ay \left( 2 C_F -  C_A + n_f T_F \right)
	+ 6  \as \ay  C_F \right] \right. \nonumber\\
& + & \left. \vphantom{\frac{1}{2}} 
	\frac{1}{2} \as^2 C_A \right) 
 +    \ay C_F \left( 
	14 \lambda_3  
	+ 40 \lambda_2 
	- 2 \lambda_1  C_A 	
	+  14 \as C_F 
	\right)	\nonumber\\
& + & \ay^2 C_F \left( 
	3  C_A - 6 C_F - 2 n_f T_F 
	\right) 
	- \frac{7}{6} \as^2 C_F C_A  	
\label{dec:s2}  \\
\dzeta{G}{2} & = & 
  \as^2 C_A^2 \left( \frac{1}{4 \varepsilon} 
  + \frac{7}{8} - \frac{1}{2} \, \Lmes \right) 
 +  \as  \ay \left( 
  \frac{2}{3} n_f T_F C_A - 2 n_f T_F C_F  
  \right) \nonumber\\ 
  & + &  \frac{7}{3} \lambda_3 \as C_A + \frac{20}{3} \lambda_2 \as C_A
  - \frac{1}{3} \lambda_1 \as C_A^2  \label{dec:G2} 
\end{eqnarray}
	and 
\begin{eqnarray}
\dzeta{m}{2} & = & \dzeta{s}{2} - \dzeta{q}{2} - \dzeta{m}{1} \dzeta{q}{1} \nonumber\\
	 & = &  
	\left( \frac{1}{\varepsilon} - 2 \Lmes \right) C_F  \left(
	\ay C_F \left[ \vphantom{\frac{1}{2}}
	3  \as  C_F 
	- \ay \left(  2 C_F - C_A +  n_f T_F \right) \right]
	 + \frac{1}{2} \as^2 C_A \right) 
	 \nonumber\\
&  + &     \ay C_F \left( 
	7 \lambda_3  
	+ 20 \lambda_2 
	- \lambda_1  C_A 	
	+  \as \left( 13 C_F - \frac{3}{2} C_A  \right)
	\right)	\nonumber\\
& + & \ay^2 C_F \left( 
	3  C_A - \frac{13}{2} C_F - \frac{3}{2} n_f T_F 
	\right) 
	- \frac{11}{12} \as^2 C_F C_A , 
	\label{dec:m2} \\
\dzeta{g_s}{2} & = & 
	\dzeta{qGq}{2} - \dzeta{q}{2} 
	- \dzeta{q}{1} \dzeta{g_s}{1} - \frac{1}{2} 
	\left( \dzeta{G}{2} + \dzeta{G}{1} \left[ \dzeta{g_s}{1} + \dzeta{q}{1} \right] 
	- \frac{1}{4} \left[ \dzeta{G}{1} \right]^2 \right)\nonumber\\
& = & 	\dzeta{cGc}{2} - \dzeta{c}{2} 
	- \dzeta{c}{1} \dzeta{g_s}{1} - \frac{1}{2} \left( 
	\dzeta{G}{2} + \dzeta{G}{1} \left[ \dzeta{g_s}{1} + \dzeta{c}{1} \right] 
	- \frac{1}{4} \left[ \dzeta{G}{1} \right]^2 \right)\nonumber\\
& = & 
\as \left(
\frac{1}{6} \lambda_1  C_A^2 
	-     \frac{7}{6} \lambda_3  C_A - \frac{10}{3} \lambda_2  C_A
	-    \ay n_f T_F \left( \frac{1}{3}  C_A -  C_F  \right) 
	-    \frac{5}{8} \as C_A^2   
   \right)
	\label{dec:gs2}.
\end{eqnarray}
	In \eqref{dec:m2} and \eqref{dec:gs2} we use perturbative expansion of \eqref{decoupling:general}.
	Clearly, \eqref{dec:m2} contains divergence and, therefore, the dependence on $\Lmes$ arises at $\mathcal{O}(\varepsilon^0)$. 
	Also there is a dependence on various $\lambda_i$. As it was noticed in Sec.~\ref{sec:decoupling_framework} bare
	decoupling relations need to be properly renormalized \eqref{decoupling:renormalized}. Let us demonstrate how this can be done at the two-loop
	level.

	Recall again that $A=\{g_s,m\}$ corresponds to high-energy theory parameters that have their counter-parts in the low-energy theory, 
	$B = \{g_y, \lambda_i\}$ denotes dimensionless parameters  and $M = \meps$ represents large masses. Let us consider perturbative
	expansion of the bare decoupling and renormalization constants that enter into \eqref{decoupling:renormalized}:
\begin{eqnarray}
		Z_A & = & 1 + \delta Z_A^{(1)}(A, B) +  \delta Z_A^{(2)} (A, B), \\
		\myunderline{Z_A} & = & 1 + \delta \myunderline{Z_A}^{(1)}(\myunderline A) +  \delta \myunderline{Z_A}^{(2)} (\myunderline A), \\
		\zeta_{A,0} & = & 1 + \dzeta{A,0}{1} (A_0,B_0,M_0) + \dzeta{A,0}{2} (A_0, B_0,M_0) \label{decoupling_constant:twoloop}.
\end{eqnarray}
	Substituting these quantities into \eqref{decoupling:renormalized} one obtains	the following expression for renormalized
	decoupling constant $\delta \zeta_A$:
\begin{eqnarray}
\dzeta{A}{1} & = & \delta Z_{A}^{(1)} (A, B) - \delta \myunderline{Z_{A}}^{(1)}( A ) + \dzeta{A,0}{1}(A,B,M),  \label{dec_const:explicit_oneloop}\\
\dzeta{A}{2} 
& = & 
  \delta Z_{A}^{(2)} (A, B) 
- \delta \myunderline{Z_{A}}^{(2)}( A ) 
- \dzeta{A}{1} \,  \left( A \, \frac{\partial}{\partial A} \right)\, \delta \myunderline{Z_{A}}^{(1)}(A )  
	\nonumber\\ 
& + & 
  \left(\delta \myunderline{Z_{A}}^{(1)}( A )  \right)^2  
- \delta Z_{A}^{(1)} (A, B) \delta \myunderline{ Z_{A}}^{(1)}( A ) 
	\nonumber \\
& + & 
   \dzeta{A,0}{1} (A,B,M) \left( \delta Z_{A}^{(1)} (A, B) - \delta \myunderline{Z_{A}}^{(1)}( A )  \right)
	\nonumber\\
& + &  \dzeta{A,0}{2}(A,B,M) 
        + \sum\limits_{x=A,B,M} \left( \delta Z_x^{(1)} \left( x \frac{\partial}{\partial x} \right) \, \dzeta{A,0}{1}(A,B,M) \right). 
                \label{dec_const:explicit_twoloop}
\end{eqnarray}
	Consequently, to find the matching relations \eqref{toy:definitions} we need to consider 
	renormalization constants of the low-energy theory, i.e., QCD in \MSbar-scheme\cite{Tarrach:1980up,Fleischer:1998dw},
\begin{eqnarray}
	Z^{\MSbar}_{m} & = &  1 - 3 C_F
			 \frac{\as}{4 \pi} 
			 \frac{1}{\varepsilon} 
		     + C_F \frac{\as^2}{(4 \pi)^2}
		     \left[
		     	\frac{1}{\varepsilon^2}
			\left(
		 \frac{11}{2} C_A
		+ \frac{9}{2} C_F
		  - 2 n_f T_F 
			\right) \right. \nonumber\\
		&& 
		\left. - \frac{1}{\varepsilon}
		\left(
		 \frac{97}{12} C_A
		+ \frac{3}{4} C_F
		-\frac{5}{3} n_f T_F 
		     \right) \right]
		    \label{ren:mMS},
		     \\	
Z^{\MSbar}_{g_s} & = & 1  +  \frac{\as}{4\pi} \frac{1}{\varepsilon}
		\left(\frac{2}{3} T_F n_f - \frac{11}{6} C_A \right) 
		+ \frac{\as^2}{(4\pi)^2}
		\left[ \frac{1}{\varepsilon^2}
		\left(
			\frac{121}{24} C_A^2 - \frac{11}{3} C_A T_F n_f
		\right.
		\right. \nonumber\\
	& & \left. \left.
		+ \frac{2}{3} n_f^2 T_F^2\right)
		+ \frac{1}{\varepsilon}
		\left( - \frac{17}{6} C_A^2 + \frac{5}{3} C_A T_F n_f
		+ C_F T_F n_f
		\right)
		\right]
		\label{ren:asMS}
\end{eqnarray}
	together with renormalization constants of the full theory, i.e., \DR\ QCD\cite{Marquard:2007uj},
\begin{eqnarray}
	Z^{\DR}_{m} & = &  1 - 3 C_F
		 \frac{\as}{4 \pi} 
		\frac{1}{\varepsilon} 
		- C_F \frac{\ay}{4\pi}
		\left( 
- 3 C_F \frac{\as}{4\pi} \frac{1}{\varepsilon}
+ \frac{\ay}{ 4\pi} \frac{1}{\varepsilon} 
		\left( 2 C_F - C_A + T_F n_f 
		\right) 
		\right) 
\nonumber\\
	&&
		     + C_F \frac{\as^2}{(4 \pi)^2}
		     \left[
		     	\frac{1}{\varepsilon^2}
			\left(
		 \frac{11}{2} C_A
		+ \frac{9}{2} C_F
		- 2 n_f T_F 
			\right) \right.\nonumber\\
	&& \left.
		- \frac{1}{\varepsilon}
		\left(
		 \frac{91}{12} C_A
		+ \frac{3}{4} C_F
		- \frac{5}{3} n_f T_F 
		\right)
		     \right] 
		    \label{ren:mDR}
\\	
Z^{\DR}_{g_y} & = & 1 - 3 C_F \frac{\as}{4\pi} \frac{1}{\varepsilon}
+ \frac{\ay}{ 4\pi} \frac{1}{\varepsilon} 
		\left( 2 C_F - C_A + T_F n_f 
		\right) \label{ren:ayDR} \\
Z^{\DR}_{\meps^2}  & = & 
	1 - 3 C_A \frac{\as}{4 \pi} \frac{1}{\varepsilon}
	+ 2 n_f T_F \frac{\ay}{4\pi}\frac{1}{\varepsilon} 
		+ \left( 7 \lambda_3 + 20 \lambda_2 - \lambda_1 C_A  \right) \frac{1}{\varepsilon}
		\label{ren:mepsDR},
\end{eqnarray}
	where we have omitted $\DR$-renormalization constant for the gauge coupling since it has
	the same form as \eqref{ren:asMS}.  Notice that for our purpose we only need one-loop renormalization for the
	evanescent coupling $\ay$ and for the mass 
	$\meps^2$.

	Given the knowledge of bare decoupling and renormalization constants in the effective and full theories, 
	one can calculate renormalized decoupling corrections. 
	Since one-loop renormalization for the gauge coupling and the quark mass 
	coincides in \DR\ and \MSbar, the first two lines in \eqref{dec_const:explicit_twoloop}
	can be represented as
\begin{eqnarray*}
	\frac{Z^{\DR}_{m}(\as,\ay)}{Z^{\MSbar}_{m}(\as^{\MSbar})} & = & 
	1 - 
	C_F C_A \frac{\as^2}{(4 \pi)^2} 
		\frac{1}{2 
		\varepsilon} 
	+ C_F C_A \frac{\as^2}{(4\pi)^2} \,\Lmes \nonumber \\
& &	 
		- C_F \frac{\ay}{4\pi}
		\left( 
- 3 C_F \frac{\as}{4\pi} \frac{1}{\varepsilon}
+ \frac{\ay}{ 4\pi} \frac{1}{\varepsilon} 
		\left( 2 C_F - C_A + T_F n_f 
		\right) 
		\right) 
		\nonumber\\
\frac{Z^{\DR}_{g}(\as,\ay)}{Z^{\MSbar}_{g}(\as^{\MSbar})} & = & 
	1  +  
	\frac{C_A}{3} \frac{\as^2}{(4\pi)^2} \left( \frac{1}{\varepsilon} - \Lmes \right)
	\left(\frac{2}{3} T_F n_f - \frac{11}{6} C_A \right) 
	. 
\end{eqnarray*}
	The renormalization of the one-loop bare decoupling constants give rise to the following:
\begin{eqnarray}
& 
& 
2\delta Z_{g_y}^{(1)} \left(\ay \frac{\partial}{\partial \ay} \right) \dzeta{m,0}{1}
	+
	\delta Z_{\meps^2}^{(1)} \left( \meps^2 \frac{\partial}{\partial \meps^2} \right) \dzeta{m,0}{1} \,  \nonumber\\
&& \qquad
	= 
2 C_F \frac{\ay}{4 \pi} 
\left[- 3 C_F \frac{\as}{4\pi} 
+ \frac{\ay}{ 4\pi} 
		\left( 2 C_F - C_A + T_F n_f 
		\right) 	
		\right]
		\left[ 
			\frac{1}{\varepsilon} - 
			\Lmes \right]  \nonumber\\
	&& \qquad
	+
	C_F \frac{\ay}{4 \pi} \left[
	\frac{\as}{4 \pi} \left( 3 C_A - 9 C_F \right)
	+ \frac{\ay}{4 \pi} \left( 6 C_F - 3 C_A  + n_f T_F\right)
	\right. \nonumber\\
	&& \left. \phantom{C_F \,\frac{\as}{(4\pi)}}
	- 7 \lambda_3 - 20 \lambda_2 + \lambda_1 C_A  \right], \\
& 
&
2 \delta Z_{g_s}^{(1)} \left( \as \frac{\partial}{\partial \as} \right)  \dzeta{g_s,0}{1}
	+
	\delta Z_{\meps}^{(1)} \left( \meps^2 \frac{\partial}{\partial \meps^2} \right) \dzeta{g_s,0}{1} \nonumber\\
&& \qquad =   
 -\frac{1}{3} C_A  \frac{\as^2}{(4 \pi)^2} 
\left[ \frac{2}{3} T_F n_f - \frac{11}{6}  C_A \right] 
 \left[ \frac{1}{\varepsilon} - \Lmes \right] \nonumber\\ 	
&& \qquad  + \frac{1}{6} C_A \frac{\as}{4 \pi} \left(  
- 3 C_A \frac{\as}{4 \pi} 
	+ 2 n_f T_F \frac{\ay}{4\pi}
		+ 7 \lambda_3 + 20 \lambda_2 - \lambda_1 C_A  
\right).
\end{eqnarray}
	In the end of the day, one gets
\begin{eqnarray}
	m^{\MSbar} 
	& = & 
	m^{\DR} 
	\left(1 + \frac{\dzeta{m}{1}}{(4\pi)} + \frac{\dzeta{m}{2}}{(4\pi)^2} \right),
	\nonumber\\
\dzeta{m}{1} & = & C_F \ay, \\
\dzeta{m}{2} & = & - \frac{11}{12} C_F C_A \as^2
		+ C_F \as \ay \left( 4 C_F + \frac{3}{2} C_A \right)
		- \frac{1}{2} C_F \ay^2 \left( C_F + n_f T_F \right),  \\
	g_s^{\MSbar} 
	& = & 
	g_s^{\DR} 
	\left(1 + \frac{\dzeta{g_s}{1}}{(4\pi)} + \frac{\dzeta{g_s}{2}}{(4\pi)^2} \right),
	\nonumber\\
	\dzeta{g_s}{1} & = & -\frac{1}{6} C_A \as, \\
\dzeta{g_s}{2} & = & - \frac{9}{8}  C_A^2 \as^2
		+ n_f T_F C_F \as \ay.
\end{eqnarray}
Obviously, the result is finite and coincides with the one that is known from literature 
(see, e.g., Ref.~\refcite{Harlander:2006rj}).
All the dependence on $\Lmes = \log \meps^2/\scale^2$ and on evanescent couplings $\lambda_i$ is canceled. Thus, by the 
explicit two-loop calculation we proved that the decoupling procedure well established 
in the context of perturbative QCD can be used not only for decoupling of heavy particles but also for $\DR \to \MSbar$ transition.

\section{The running mass of the b-quark: Matching QCD and SUSY QCD}\label{sec:qcd-sqcd}
	In this section, we consider the SUSY QCD part of the MSSM as a full theory.
	The Lagrangian of SUSY QCD can be found, e.g.,  in Ref.~\refcite{Bednyakov:2002sf}. 
	After dimensional reduction in addition to \eqref{lag:qcd-escalar} there arise 
	interactions of $\varepsilon$-scalars with squarks and gluinos(see~\ref{sec:escalar-sqcd}).
	The task again is to find a relation 
\begin{equation}
	\mbms(\scale) = \mbdr(\scale) 
	\times \zeta_{m_b} \left(\as^{\DR}, M^{\DR}, \scale\right),
\end{equation}
	where $M^{\DR}$ corresponds to the \DR-renormalized masses of heavy particles. In the
	considered case
\begin{equation}
	M = \{ m_t, m_{\tilde g}, m_{\tilde q_i} \},\quad q = \{u,d,c,s,t,b\},\,i = 1,2,
\end{equation}
	where $m_{\tilde g}$ denotes the gluino mass and $m_{\tilde q_i}$ corresponds to squark masses.

	Let us begin with the one-loop result for the mass and the gauge coupling (see, e.g., Ref~\refcite{Pierce:1996zz}):
\begin{eqnarray}
	\delta \zeta^{(1)}_{g_s} & = & 
	\frac{\as}{4 \pi} \left( \frac{1}{3} \log \frac{m_t^2}{\scale^2} 
	+ \frac{C_A}{3} \log\frac{m_{\tilde g}^2}{\scale^2}
	+ \frac{1}{12} \sum\limits_{q}  \sum\limits_{i=1}^{2} \log \frac{m_{\tilde q_i}^2}{\scale^2} 
	- \frac{C_A}{6}  \right), 
	\label{dzeta:gs_finite} \\
\delta \zeta^{(1)}_{m_b} & = & \frac{\as}{ 8 \pi } C_F  
	\left( 
	1 
	+ \frac{m^2_{\tilde g}}{m^2_{\tilde b_1} - m^2_{\tilde g}} 
	+ \frac{m^2_{\tilde g}}{m^2_{\tilde b_2} - m^2_{\tilde g}} 
	+ 2 \log \frac{m^2_{\tilde g}}{\scale^2} \right.\nonumber\\
	& + &  \log \frac{m^2_{\tilde b_1}}{m^2_{\tilde g}}  
		\left( 1 
		- \frac{m^4_{\tilde g}}{(m^2_{\tilde b_1} - m^2_{\tilde g})^2} 
		- 2 \sin 2 \theta_b \frac{m_{\tilde g}}{m_b} 
			\frac{m^2_{\tilde b_1}}{m^2_{\tilde b_1} - m^2_{\tilde g}} 
			\right) \nonumber\\
	& + &  \log \frac{m^2_{\tilde b_2}}{m^2_{\tilde g}}  
		\left( 1 
		- \frac{m^4_{\tilde g}}{(m^2_{\tilde b_2} - m^2_{\tilde g})^2} 
		+ 2 \sin 2 \theta_b \frac{m_{\tilde g}}{m_b} 
			\frac{m^2_{\tilde b_2}}{m^2_{\tilde b_2} - m^2_{\tilde g}} 
			\right) \label{dzeta:mb_finite}. 
\end{eqnarray}
Here $\theta_b$ is the bottom squark mixing angle.  
Unphysical $\varepsilon$-scalars contribute 
$-\frac{\as}{4 \pi}\frac{C_A}{6}$ to the gauge 
decoupling constant and $\frac{\as}{4 \pi} C_F$ to the quark mass 
decoupling constant.  It should be noted that \eqref{dzeta:mb_finite} is nothing
	else but the one-loop contribution to the pole
	mass of the quark from superparticles. 
	
	One may notice
	the dangerous dependence on  $m_b$ in 
	\eqref{dzeta:mb_finite}. As it was stated in 
	Sec.~\ref{sec:decoupling_framework}, decoupling constants
	should not depend on low mass scales. This contradiction 
	is due to the fact that in the MSSM quarks acquire 
	their masses after spontaneous breakdown of the electroweak
	symmetry (SSB)  . 
	In spite of the fact that we neglect interactions
	parametrized by Yukawa couplings they obviously manifest themselves
	in quark masses.  
	Due to the supersymmetry squark interactions
	with Higgs bosons are also parametrized by the same
	Yukawa couplings. 
	After SSB squark quadratic Lagrangian receives 
	a contribution proportional to the mass of a quark $m_q$
\begin{equation}
	\delta \L_{\tilde q \tilde q} 
	= - m_q^2 
	\left( \tilde q^*_{L} \tilde q_{L} +  
	\tilde q^*_{R} \tilde q_{R} \right)  
	- m_q a_q  
	\left( \tilde q^*_{L} \tilde q_{R} +  
	\tilde q^*_{R} \tilde q_{L} \right),~
	a_q = A_q - \bar \mu \left\{ \cot \beta,\,\tan \beta \right\},
	\label{squark_mixing}
\end{equation}
	where $\tilde q_{L,R}$ correspond to squark fields
	, and $A_q, \bar \mu$, $\tan\beta$ are the MSSM
	parameters\footnote{We use $\bar \mu$ to denote supersymmetric Higgs mixing parameter in order to distinguish it from 
	the renormalization scale $\scale$}.
	In the definition of $a_q$ 
	for up-squarks one has to choose $\cot\beta$
	and for down-squarks --- $\tan\beta$.
	Usually one considers \eqref{squark_mixing} as an additional
	contribution to the squark mass matrix and 
	after diagonalization introduces mass eigenstates $m^2_{{\tilde q}_1}$, 
	$m^2_{{\tilde q}_2}$ and a mixing angle $\theta_q$
	that implicitly depend on $m_q$. 
	If one takes into account that 
\begin{equation}
	\sin 2 \theta_q  =  
	\frac{2 m_q a_q}{m^2_{ {\tilde q}_1} - m^2_{ {\tilde q}_2}}
	\label{squark:mixing}
\end{equation}
it is possible to cancel dangerous powers of $m_b$ in \eqref{dzeta:mb_finite}.  	
	However, it is not the end of the story, since
	the mass eigenvalues also depend on $m_q$. 

\begin{figure}[h]
	\begin{center}
		\includegraphics{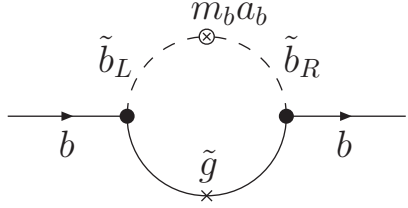}
	\end{center}
	\caption{Feynman diagram with one insertion of $m_b a_b$ that
	contributes to the self-energy of the quark at the leading
	order in $m_b/M$}
	\label{fig:squark_mixing}
\end{figure}

	There are two equivalent ways to obtain decoupling
	corrections that are formally independent of low mass scales.
	The first one is to reexpand \eqref{dzeta:mb_finite}
	and \eqref{dzeta:gs_finite} in $m_b/M$. The second one
	is to consider \eqref{squark_mixing} as 
	a part of the interaction Lagrangian from the very beginning. 
	Clearly, insertion of the vertices from \eqref{squark_mixing} in 
	a Feynman diagram gives rise to a contribution that is 
	proportional to some power of $m_b$. 
	In the context of the asymptotic expansion 
	only a finite number of these insertions 
	has to be taken into account.
	For example, if one considers quark self-energy and 
	the leading terms in the asymptotic expansion, it
	is sufficient to take into account only one insertion that
	mixes ``left-handed'' and ``right-handed'' squarks 
	(see Fig.~\ref{fig:squark_mixing}). 
	This approach allows one to keep all the dependence 
	on $m_b$ explicit and obtain decoupling constants
	that are independent of $m_b$ and exhibit  
	perfect factorization property. 
	Nevertheless, the dependence on $m_b$ of
	\eqref{dzeta:gs_finite} and \eqref{dzeta:mb_finite} is
	analytic as $m_b \to 0$ and formally these expressions
	differ from the perfectly factorized ones only by
	the terms $\mathcal{O}(m_b^2/M^2)$ that are negligible even for 
	$M \sim 0.1~\TeV$. 
	In our work we decided to keep the answer in the form 
	that can be obtained from \eqref{dzeta:mb_finite} by substitution
	of \eqref{squark:mixing}. This trick also works at the two-loop level. 

	The evaluation of the two-loop decoupling constant for the 
	$b$-quark mass goes along the same lines as in the previous section.
	The important thing that has to be mentioned is that in SUSY QCD
	$\delta \zeta_{q_L} \neq \delta \zeta_{q_R}$.
	In this case\footnote{we suppress the dependence
	of the self-energy on large mass scales},
\begin{eqnarray}
i \Gamma_{q} (\hat p,m) & = & 
	\Sigma_L(p^2,m^2)
	\, \hat p P_L
	+ \Sigma_R(p^2,m^2)
	\, \hat p P_R
	+ \Sigma_s(p^2,m^2) 
	\, m, \quad P_{L,R} = \frac{1 \mp \gamma_5}{2}, \nonumber\\
\delta \zeta_{q_l} & = & 
	\left. \left(-i  \frac{1}{2 \, n_c \, p^2} \times 
	\Tr  \, P_l \, \hat p \, \Gamma_{q} \, \right)\right|_{p,m=0}
	= - \Sigma_l(0,0), \quad l=L,R 
\label{bare_dec:q_lr}
\end{eqnarray}
	and 
\begin{align}
\dzeta{m}{1}  =  
	\dzeta{s}{1} 
	- & \frac{1}{2} \left( \dzeta{q_L}{1} + \dzeta{q_R}{1} \right), 
	\label{dec:m1_lR} \\
\dzeta{m}{2}  =  
	\dzeta{s}{2} 
	- & \frac{1}{2} \left( \dzeta{q_L}{2} + \dzeta{q_R}{2} 
	+ \dzeta{m}{1} \left[ \dzeta{q_L}{1} + \dzeta{q_R}{1} \right] \right.\nonumber\\
	- & \left. \frac{1}{4} \left[ \dzeta{q_L}{1} - \dzeta{q_R}{1} \right]^2
	\right). 
\end{align}
	Since only quarks with the same chirality enter in the quark-gluon vertex,  
	the expression for the gauge coupling decoupling constant (see \eqref{dec:gs2})
	is modified in a straightforward manner.

	The calculation is performed in the \DR-scheme with an explicit mass term for 
	the $\varepsilon$-scalars. 
	Almost all needed renormalization constants for SUSY QCD can be found in
	Ref.~\refcite{Bednyakov:2002sf}. The only exception is the $\meps^2$ counter-term
	that looks like
\begin{equation}  \label{ren:meps_ct_susy}
Z_{\meps^2}  = 1 +  \frac{\as}{4\pi} \frac{1}{\varepsilon}  
	\left[ 2 n_f T_F - 3 C_A + 
	\frac{1}{\meps^2} 
	\left(
	2 T_F \, \sum\limits_{n_{\tilde f}} m^2_{\tilde f} 
	- 4 T_f \, \sum\limits_{n_f} m_f^2
	- 2 C_A m^2_{\tilde g} 
	\right)
	\right],
\end{equation}
	where $n_{\tilde f}$ is used to denote the sum over different squarks
	and $n_f$ corresponds to the summation over quark flavours.
	Notice that when SUSY is not broken by soft terms, $m_{\tilde g}^2 = 0$
	and $m_q^2 = m_{\tilde q_i}^2$, so the renormalization constant \eqref{ren:meps_ct_susy}
	and, thus, beta-function is homogeneous with respect to $\meps^2$
	and one can safely put $\meps^2=0$ from the very beginning.

	As it was mentioned in Sec.~\ref{sec:decoupling_framework}, it is possible 
	to obtain the same expression 
	by considering some observable
	that can be defined in both the effective and high-energy theories. 
	For example, one can use the pole mass $M_b$ as an intermediate quantity to
	find the relation between $\mbms$ and $\mbdr$. Since the 
	two-loop SUSY QCD expression for $M_b$ has been found earlier\cite{Bednyakov:2002sf},
	it is easy to calculate $\dzeta{m_b}{2}$ given the knowledge
	of the one-loop decoupling constants \eqref{dzeta:gs_finite}--\eqref{dzeta:mb_finite}.
	Let us briefly describe this approach, since we use it to cross-check our result. 
	
	Consider the two-loop relation between $M_b$ and $\mbms$ calculated within the QCD\cite{Tarrach:1980up}
\begin{equation} \label{pole_mass:QCD}
	M^{\MSbar}_b = m_b \left[1 + \sigma^{(1)}(m_b,\as) +  \sigma^{(2)}(m_b,\as)\right],
	\quad m_b \equiv \mbms, \as \equiv \as^{\MSbar}.\
\end{equation}
	One can rewrite \eqref{pole_mass:QCD} in terms
of SUSY QCD \DR-parameters by means of decoupling constants
($m_b \equiv \mbdr, g_s \equiv g_s^{\DR}$)
\begin{align} \label{pole_mass:QCD_with_dc}
	M_b^{\MSbar}  =  m_b \Bigl[1 &+ \left( \sigma^{(1)} + \dzeta{m_b}{1} \right) 
	+ \left( \sigma^{(2)} + \dzeta{m_b}{2} \right) 	\nonumber\\	
	&+ 
	 \dzeta{m_b}{1} \Bigl(1 + m_b \frac{\partial}{\partial m_b} \Bigr) \sigma^{(1)} 
	+ 2 \dzeta{g_s}{1} \Bigl( \as \frac{\partial}{\partial \as} \Bigr)  \sigma^{(1)}
	 \Bigr], 
\end{align}
	where $\sigma^{(1,2)}$ are the same functions of their arguments as
	in \eqref{pole_mass:QCD}, i.e., they correspond to the diagrams with
	quarks and gluons only. As it was stated in 
	Sec.~\ref{sec:decoupling_framework}, expression \eqref{pole_mass:QCD_with_dc}
	allows one to approximate the result of the full theory. 
	If $M_b^{\DR}=M^{\DR}_b(\mbdr, M^{\DR}, \dots)$ is
	the pole mass calculated within SUSY QCD, at the leading order of \LME\ 
	we have $M_b^{\MSbar} = M_b^{\DR}$ and 
\begin{align}
	\dzeta{m_b}{2}   =  
	\frac{M^{\DR}_b - m_b }{m_b}  - & 
	\left(  \sigma_1 + \dzeta{m_b}{1} \right)  \nonumber\\
	  - & \left(\sigma_2 
	   +  
	 \dzeta{m_b}{1} \Bigl(1 + m_b \frac{\partial}{\partial m_b} \Bigr) \sigma^{(1)} 
	+ 2 \dzeta{g_s}{1} \Bigl( \as \frac{\partial}{\partial \as} \Bigr)  \sigma^{(1)}
	\right). 
	\label{mass:2loop}
\end{align}

	Direct application of \eqref{mass:2loop} gives rise to  
	analytic expression for the $m_b$ decoupling constant that is free
	from $\log m_b/\scale$ but differs 
	from the one obtained by the procedure described earlier. 
	Careful investigation of the difference shows 
	us that the discrepancy is due to the fact 
	that both the results lack the perfect factorization property. 
	The difference appears to be proportional to 
	some power of $\sin 2\theta_b$ and formally can 
	be rewritten in such a way that it will 
	be $\mathcal{O}(m_b^2/M^2)$. 
	Indeed, numerical analysis shows
	that additional terms in the considered regions
	of the MSSM parameter space 
	amount to $10^{-3}$ \% shift in the result.  

	The calculation of the corrections is carried out by means of a computer 
	program written in FORM\cite{Vermaseren:2000nd}. 
	Two-loop bubble integrals that appear in \LME\ 
	are recursively reduced to a master-integral\cite{Davydychev:1992mt}
	by integration by the parts method\cite{Chetyrkin:1981qh}. The numerical
	evaluation of the master integral is carried out with the help of 
	C++ library bubblesII\cite{bubblesII}. 

\section{Results}
	Since we decouple all the heavy particles at the same time, 
	this results in the huge expressions for the decoupling constants that depend on all the heavy mass scales
	of the model. Consequently, we will not present the answer in great detail as in Sec.~\ref{sec:qcd-qcd}
	but just give a numerical impact of the result. 

	Evaluation of the corrections requires the knowledge of running MSSM parameters.
	However, the precise values are unknown, so one usually uses some hypothesis to reduce the 
	parameter space of the model. The main uncertainty comes from the unknown soft terms.
	To reduce the number of free parameters, the so-called universality hypothesis is
	usually adopted, i.e., one assumes the universality or equality of various soft parameters
	at high energy scales. With the universality hypothesis one is left with the following set of free 
(mSUGRA\cite{Chamseddine:1982jx,Barbieri:1982eh,Hall:1983iz,Nath:1983aw}
      )	
	parameters:
$$m_0,~m_{1/2},~A_0~\mathrm{and}~\tan\beta = \frac{v_2}{v_1}.$$
	Here $m_0$, $m_{1/2}$ are universal scalar and fermion masses.
	They define mass splitting between the SM particles and their superpartners.
	Soft cubic interactions are parametrized by $A_0$ and $\tan\beta$
	is the ratio of vacuum expectation values of the Higgs fields. Also the sign of
	$\bar \mu$ is not fixed. In what follows we assume that $\bar\mu>0$.

	Usually, some computer code\cite{Allanach:2001kg,Porod:2003um,Djouadi:2002ze,Paige:2003mg} is used 
	to take an advantage of renormalization group method and 
	calculate spectra and other observables. The universal boundary conditions are 
	applied at some high energy scale \Mgut. 
	However, it is inconvenient to calculate low-energy observables 
	in terms of parameters defined at \Mgut. 
	One has to use the renormalization group to obtain the values of 
	the corresponding parameters at the electroweak scale \Mew\ which is of our interest. 
	There arises another complication, since for running one needs to know the values
	of dimensionless couplings at \Mgut. 
	In contrast to soft terms gauge and Yukawa
	couplings are severely constrained by known electroweak physics, so
	natural boundary conditions for them are defined at \Mew. For most
	of the SM parameters these conditions are nothing else but 
	relations of the type discussed in this work \eqref{decouling_constants:definition}, so 
	they are functions of (almost) \emph{all} the parameters of the MSSM.
	To break this vicious circle, one usually makes a (reasonable) initial guess 
	for unknown parameters either at \Mgut\ or at \Mew\ and 
	after some iterations a stable solution for the equations is obtained. 

\begin{figure}[h]
	\begin{center}
		\includegraphics[scale=0.9]{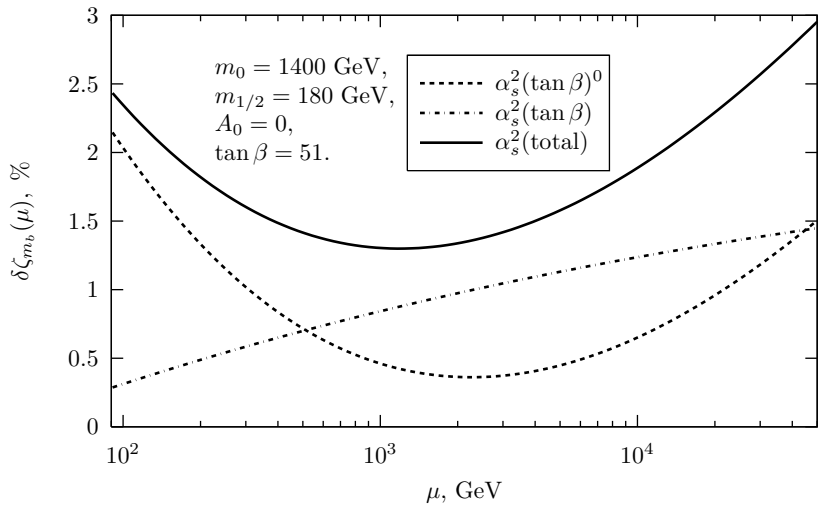}
	\end{center}
	\caption{The dependence of two-loop decoupling corrections $\dzeta{m_b}{2}$ on the scale \scale.
	Lines marked by $\as^2 (\tan\beta)$ correspond to the 
	contribution that is proportional to $\tan \beta$. Lines labeled by $\as^2 (\tan\beta)^0$ correspond to
	terms that lack such dependence on $\tan \beta$. Terms with $\tan^n\beta$, $n>1$ turn out to be suppressed}
	\label{fig:two-loop-running}
\end{figure}

	In order to demonstrate our result, we present the values of two-loop corrections
	evaluated with running parameters given by the SOFTSUSY code\cite{Allanach:2001kg}.
	The decoupling constant for $m_b$ explicitly depends on the scale \scale. 
	In order to reduce the uncertainty associated with large logarithms, one has to 
	choose $\scale \sim M$. Indeed, Fig.~\ref{fig:two-loop-running} shows the dependence of
	two-loop corrections on the scale $\scale$ for the specific point 
	preferred by combined EGRET\&WMAP constraints\cite{deBoer:2005bd}. 
	One sees that for $\scale \sim 1~\TeV$ the calculated correction is about 1.5 \%.
	
	Figure~\ref{fig:two-loop-running} also addresses another issue related to contributions
	that can be potentially enhanced by large $\tan \beta$. 
	Since in our approach $\tan \beta$ appears only through mixing \eqref{squark:mixing}, it
	is easy to trace this dependence. Clearly, only first power of $\tan\beta$ should be taken 
	into account at the leading order of $m_b/M$ expansion. 
	From Fig.~\ref{fig:two-loop-running} one
	sees that even for large $\tan\beta \simeq 51$ corrections 
	$\propto \tan\beta$ do not give a dominant contribution, so one should keep
	other terms in a careful analysis.

	In the above-mentioned computer codes the relation between $\mbdr(\scale)$ and $\mbms(\scale)$  is
	usually used at $\mu=\Mew$. In what follows we also employ this choice for matching. 
	However, one should keep in mind the it is not the optimal scale for $\dzeta{m_b}{2}$ evaluation.

\begin{figure}[h]
		\begin{tabular}{cc}
			\includegraphics[scale=0.8]{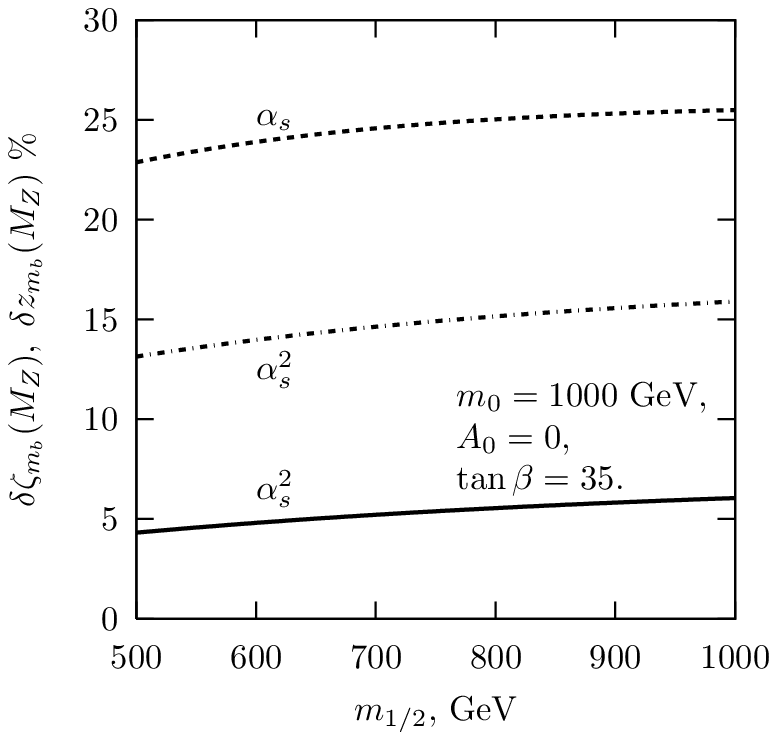} & 
			\includegraphics[scale=0.8]{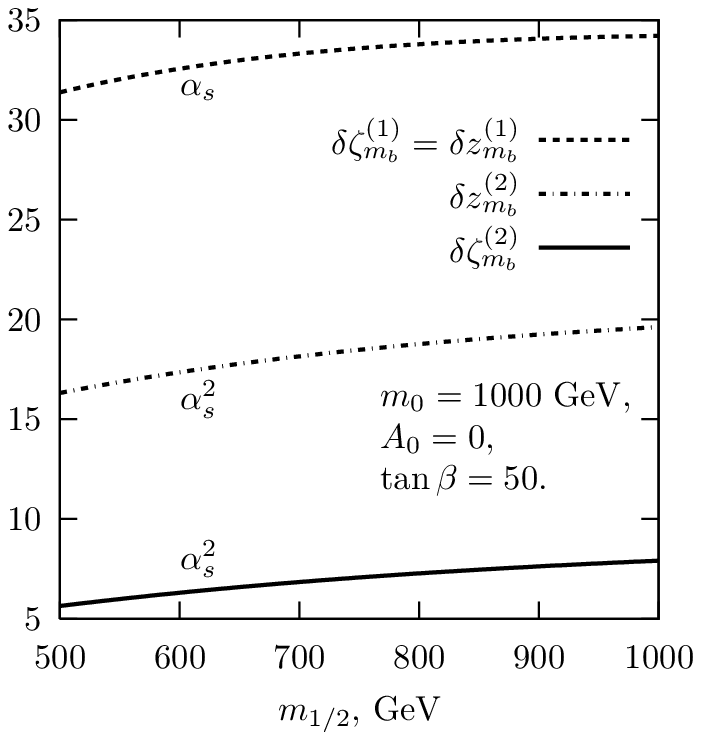}  
		\end{tabular}
		\caption{Different SUSY QCD corrections to the $b$-quark 
		pole mass $M_b$ and the decoupling constant $\zeta_{m_b}$ as functions of $m_{1/2}$.
		Here $\delta z_{m_b} \equiv (M_b - m^{\DR}_b)/m^{\DR}_b$
		and $\delta \zeta_{m_b} \equiv (m^{\MSbar}_b - m^{\DR}_b)/m^{\DR}_b$.	
		At the leading order of Large Mass Expansion and at the one-loop level $\dzeta{m_b}{1} = \delta z_{m_b}^{(1)}$.
		However, for two-loop corrections $\dzeta{m_b}{2} \neq \delta z_{m_b}^{(2)}$.
		}
		\label{fig:dzeta_vs_pm}
	\end{figure}

\begin{figure}[h!]
		\begin{center}
			\includegraphics[scale=0.95]{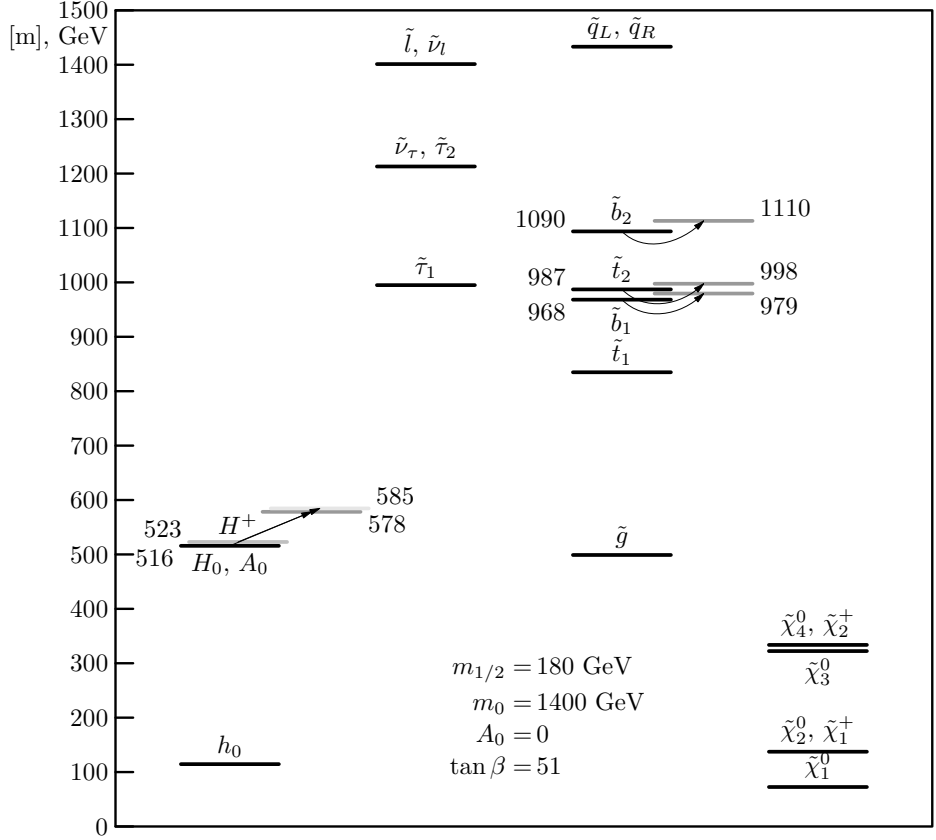}  
		\end{center}
		\caption{Superparticle spectrum for the so-called EGRET\&WMAP point of the MSSM 
		parameter space. The shifts in mass values
		due to two-loop $b$-quark decoupling corrections are also presented (shifts less than one per cent are not
		shown).}
		\label{fig:spectrum_egret}
	\end{figure}

	The final aim of the calculation is to insert calculated correction
	to the $m_b$ decoupling constant into the above-mentioned iterative process. 
	We stress again that contrary to the $t$-quark case\cite{Bednyakov:2005kt} 
	the SUSY QCD contribution to the $b$-quark pole mass\cite{Bednyakov:2002sf} 
	should not be directly applied to the calculation of $\mbdr$.
	In Ref.~\refcite{Bednyakov:2005kt}, 
	the two-loop SUSY QCD result for $M_b$ was implicitly used   
	as an estimate of the decoupling correction.  
	At the one-loop level this is reasonable but it is not true at higher loops. 
	Figure~\ref{fig:dzeta_vs_pm} shows a typical dependence of the corrections to $\delta\zeta_{m_b}$ 
	on $m_{1/2}$ for certain values of other parameters of the model. 
	For comparison we also plot pole mass corrections $\delta z_{m_b} \equiv (M_b - m^{\DR}_b)/m^{\DR}_b$. 
	It is clear that in the analysis of Ref.~\refcite{Bednyakov:2002sf}, 
	$\delta z_{m_b}$ overestimates $\delta \zeta_{m_b}$.
	Nevertheless, it was demonstrated\cite{Bednyakov:2002sf} 
	that for a wide region of parameter space even overestimated SUSY QCD corrections 
	do not influence superparticle spectrum significantly. 
	They only become important for large values
	of $\tan\beta$, since in this case $b$-quark Yukawa coupling obtained from the running mass is 
	also large. Indeed, Fig.~\ref{fig:spectrum_egret} shows superparticle spectra for the EGRET\&WMAP\cite{deBoer:2005bd} point 
	obtained by SOFTSUSY together with the shifts for the masses 
	after inclusion of our result in the code. One sees that for large $\tan\beta$ two-loop corrections mostly influences 
	a heavy Higgs spectrum\cite{Allanach:2003jw}. 

\section{Conclusions}
	The mass parameter of the $b$-quark plays an important role in phenomenological analysis
	of the MSSM. Strong interactions usually give rise to large radiative corrections to the quark mass
	and, thus, have to be calculated and taken into account. 
	In this work we have proposed a method
	that allows one to find the value of the SUSY QCD \DR-running $b$-quark mass $m^{\DR}_b$ 
	directly from the corresponding
	value of \MSbar-mass $\mbms$ defined in the QCD. We consider the QCD as 
	the low-energy effective theory of the more fundamental SUSY QCD and obtain 
	the relation between $\mbms$ and $\mbdr$ by decoupling of heavy particles.

	The transition from \DR\ to \MSbar\ scheme can be achieved almost
	automatically by decoupling of unphysical $\varepsilon$-scalars together with physical squarks and gluinos.
	To justify the latter statement, decoupling of $\varepsilon$-scalars is considered in
	the context of \DR\ QCD and known relations between \DR- and \MSbar-parameters are obtained.

	Applying a general matching procedure to the SUSY QCD case we calculate a two-loop contribution to
	the decoupling constant $\zeta_{m_b}$ for the $b$-quark running mass. This in turn allows one 
	to determine $\mbdr$ more precisely from known SM input and implement three-loop
	running of the MSSM parameters (see~Ref.~\refcite{Jack:2004ch}) consistently.
	The numerical analysis of
	the correction and its impact on the spectrum is carried out. 
	One, however, should keep in mind, that for the $b$-quark Yukawa interactions neglected in SUSY QCD 
	give a sizable contribution to the pole mass\cite{Bednyakov:2004gr}. Having in mind \eqref{mass:2loop}, 
	one may try to calculate corrections to $\zeta_{m_b}$ from the decoupling of Higgs bosons and their
	superpartners. We will study this issue elsewhere. 
	
	Finally, let us stress again the advantages and disadvantages of the proposed method.	
	The main advantage seems obvious. One need not to consider evanescent couplings and 
	their renormalization in nonsupersymmetric theories as, e.g., in 
	Refs.~\refcite{Harlander:2005wm},\refcite{Harlander:2007wh}. 
	However, one has to pay some price 
	for this simplification, since separate treatment of massive $\varepsilon$-scalars is required.
	For our problem we implemented the corresponding Feynman rules in FeynArts 
	package
	and generated needed diagrams by computer.

	Another obvious issue is a simultaneous  decoupling of all heavy particles. This is reasonable
	only if the corresponding masses are of the same order, which may be not true for some SUSY scenarios, e.g. for 
	Split SUSY\cite{Giudice:2004tc}. In the latter case, a step-by-step decoupling is needed. 
	Nevertheless, a $\DR\to\MSbar$ transition is required at some stage and we think that this 
	step can be carried out by decoupling of $\varepsilon$-scalars. It is reasonable to do this as soon 
	as possible, since in this case no evanescent couplings appear in the effective nonsupersymmetric theory.

\section{Acknowledgements}
\vspace{5mm}
	The author would like to thank D.I.~Kazakov, A.~Sheplyakov, and T.~Hahn for fruitful discussions. 	
	Financial support from the Russian Foundation for Basic Research (grant \# 05-02-17603) is kindly acknowledged.
\appendix
\section{The $\varepsilon$-scalars in the QCD}\label{sec:escalar-qcd}
	First of all, consider pure gauge QCD Lagrangian in four dimensions
\begin{eqnarray*}
	{\cal L}_{QCD} & = & 
		- \frac{1}{4} F^a_{\mu \nu}F_a^{\mu \nu} \\
	F^a_{\mu\nu} & = & \partial_{\mu} G^a_{\nu} - 
			   \partial_{\nu} G^a_{\mu} -
			   g_s f^{abc} G^b_{\mu} G^c_{\nu}
\end{eqnarray*}
	Performing Dimensional Reduction from space-time dimension four to 
	$d=4 - 2 \varepsilon$ we should split four-vector into $d$-vector 
	and so-called $\varepsilon$-scalars
\begin{eqnarray*}
		 d = 4  & \rightarrow  & d = 4 - 2 \varepsilon \\
	       	\mu & \rightarrow &  \left(\mu, \hat\mu\right) \\
		g_{\mu\nu}^4 & \rightarrow & \left( g_{\mu\nu}^{4 - 2\varepsilon},
						    g_{\hat\mu\hat\nu}^{2 \varepsilon} 
						\right) \\
		G^a_{\mu} & \rightarrow & \left(G^a_{\mu}, G^a_{\hat\mu} \right)
\end{eqnarray*}
	``Coordinates'' that correspond to $2\varepsilon$ sub-space are assumed to be space-like, so 
	$g^{2 \varepsilon}_{\hat\mu \hat\mu} = -1$ (no summation). 
	In what follows we use Latin letters to denote $2 \varepsilon$
	scalar indices and Greek letters for the Lorentz ones, i.e.,
\begin{equation}
	g_{\hat \mu \hat \nu}^{2\varepsilon} \rightarrow g_{ij}, 
\qquad  g_{\mu \nu}^{4-2\varepsilon} \rightarrow g_{\mu\nu}, 
\qquad  G^a_{\hat\mu} \rightarrow W^a_i, 
\qquad  G_a^{\hat\mu} \rightarrow g^{ij} W^a_j = -W^a_i.
\end{equation}
	Since all the fields after dimensional reduction 
	do not depend on $2 \varepsilon$ coordinates, the corresponding 
	derivatives (momenta) are zero. Consequently, 
\begin{eqnarray*}
	F^a_{\mu\nu} F_a^{\nu\nu} & \rightarrow & 
		F^a_{\mu\nu}F_a^{\mu\nu} + 
		F^a_{\mu\hat\nu} F_a^{\mu\hat\nu} + 
		F^a_{\hat \mu\nu} F_a^{\hat\mu\nu} + 
		F^a_{\hat\mu\hat \nu}F_a^{\hat\mu\hat\nu}  \\
	F^a_{\mu\hat\nu} & = & 
		\partial_{\mu} G^a_{\hat\nu} - g_s f^{abc} G^b_{\mu}G^c_{\hat\nu} \\
	F^a_{\hat \mu\nu} & = & 
		- \partial_{\nu} G^a_{\hat\mu} - g_s f^{abc} G^b_{\hat \mu}G^c_{\nu} = - F^a_{\nu\hat\mu} \\
	F^a_{\hat \mu\hat\nu} & = & - g_s f^{abc}  G^b_{\hat\mu} G^c_{\hat\nu} \\
	F^a_{\mu\hat\nu} F_a^{\mu\hat\nu} + 
	F^a_{\hat \mu\nu} F_a^{\hat\mu\nu} & = &  
	2 F^a_{\mu\hat\nu} F_a^{\mu\hat\nu} \\
	-\frac{1}{4} 2 F^a_{\mu\hat\nu} F_a^{\mu\hat\nu} 
	& = &  -\frac{g^{ij}}{2} 
		 \partial_{\mu} W^a_i \partial^{\mu} W^a_j
		 - \frac{g^{ij}}{2} g_s^2 f^{abc} f^{a\tilde b \tilde c} 
		G^b_{\mu} W^c_{i} G_{\tilde b}^{\mu} W^{\tilde c}_{j} \\
	& + & 
\frac{g^{ij}}{2} g_s f^{abc} G_b^{\mu}
		 	\left(
				( \partial_{\mu} W^a_i ) W^c_{j}  
			-	( \partial_{\mu} W^c_i ) W^a_{j}   
			\right)
\end{eqnarray*}
	and the Lagrangian of pure gauge QCD after dimensional reduction looks like
\begin{align}
	{\cal L}^{4 - 2 \varepsilon}_{QCD}  =  
		- & \frac{1}{4} F^a_{\mu\nu} F_a^{\mu\nu}
		- \frac{g^{ij}}{2} \partial_{\mu} W^a_i \partial^{\mu} W^a_j
		+ g^{ij} g_s f^{abc} \partial_{\mu} W^a_i G_b^{\mu} W^c_j \nonumber\\
		 - &  \frac{g^{ij}}{2} g_s^2 f^{abc} f^{a\tilde b \tilde c}  
		G^b_{\mu} W^c_{i} G_{\tilde b}^{\mu} W^{\tilde c}_{i} 
		- \frac{g^{ij} g^{kl}}{4} 
		g_s^2 f^{abc} f^{a\tilde b \tilde c}    
		W^b_i W^c_k W^{\tilde b}_j W^{\tilde c}_l \nonumber\\
	 =  
		- & \frac{1}{4} F^a_{\mu\nu} F_a^{\mu\nu}
		- \frac{g^{ij}}{2} 
		\left(\partial_\mu \delta^{ac}
			- g_s f^{abc} G^{b}_{\mu} 
		\right) W^c_i 
		\left(\partial^\mu \delta^{a\bar c}
			- g_s f^{a\bar b \bar c} G_{\bar b}^{\mu} 
		\right) W^{\bar c}_j \nonumber\\
	 - & 
	\frac{g^{ij} g^{kl}}{4} 
		    g_s^2 f^{abc} f^{a\tilde b \tilde c}    
		W^b_i W^c_k W^{\tilde b}_j W^{\tilde c}_l \nonumber\\
	 =  
		- & \frac{1}{4} F^a_{\mu\nu} F_a^{\mu\nu}
		- \frac{g^{ij}}{2} 
		 \left(D_\mu W_i\right)_a  
		 \left(D^\mu W_j\right)_a  
		- 
	\frac{g^{ij} g^{kl}}{4} 
		    g_s^2 f^{abc} f^{a\tilde b \tilde c}    
		W^b_i W^c_k W^{\tilde b}_j W^{\tilde c}_l,
		\label{lag:pureQCD}
\end{align}
	where we introduced a covariant derivative 
\begin{eqnarray*}
	\left( D_{\mu} \right)_{ij} & = &
	\partial_{\mu} \delta_{ij} + i g_s T^a_{ij} G^a_{\mu}. 
\end{eqnarray*}
	Here $T^a_{ij}$ - generator of a gauge group in some representation. 
	For the $\varepsilon$-scalars we have 
\begin{equation}
	D^{ac}_{\mu} = \partial_{\mu} \delta^{ac}
	+ i g_s (- i f^{bac} ) G^{b}_{\mu} 
		\label{derivative:adjoint}.
\end{equation}
	Consider gauge transformations of the fields with infinitesimal
	parameter $\omega^a$
\begin{eqnarray}
	\delta G^a_{\mu} & = & \partial_{\mu} \omega^a
	- g_s f^{abc} G^b_{\mu} \omega^c \\
	\delta W^a_{i} & = & - g_s f^{abc} W^b_{i} \omega^c
\end{eqnarray}
	All three terms in 
	\eqref{lag:pureQCD} are invariant under these transformations   
	separately. As a consequence, couplings of 
	gluon-gluon-$\varepsilon$-scalar and 
	gluon-gluon-$\varepsilon$-scalar-$\varepsilon$-scalar vertices
	are fixed by gauge invariance to be equal to $g_s$. 
	On the contrary, gauge transformations do not mix $\varepsilon$-scalar
	four-vertex with something else.

	The $\varepsilon$-scalar part of the action in momentum representation 
	looks like
\begin{eqnarray*}
	S_{\varepsilon} & = & \int d^d p_1 
		\left[
			\frac{p_1^2}{2} W^a_i (p_1) 
			W^a_i(-p_1) 
		\right] \\
			& - & 
			i g^{ij}
		\frac{g_s}{2} f^{abc} \int d^dp_1 d^d p_2 
		\left(p_2^{\mu} - p_1^{\mu}
		\right) W^a_i(p_1) W^b_j(p_2) 
		G^c_{\mu} (-p_1-p_2) \\ 
			& - & 
			 g^{ij}
	 \frac{g_s^2}{2} f^{abc} f^{ade} \int d^dp_1 d^d p_2d^dp_3 
		 W^c_i(p_1) W^e_j(p_2) G^b_{\mu} (p_3) 
		 G_d^{\mu} (-p_1-p_2-p_3) \\ 
& - & 
	g^{ij}
	g^{kl}
	\frac{g_s^2}{4} f^{abc} f^{ade} \int d^dp_1 d^d p_2d^dp_3
		 W^b_i(p_1) W^c_k (p_2) 
		 W^d_j  (p_3) 
		 W^e_l(-p_1 - p_2 - p_3).
\end{eqnarray*}
	The corresponding Feynman rules can be derived from the action by taking a functional
	derivative with respect to the fields
\begin{eqnarray}
	i \frac{\delta^3 S}{\delta W^{a}_{i} (k_1) 
			    \delta W^{b}_{j} (k_2) 
			    \delta G^{c}_{\mu} (k_3)} & = & 
	g_s f^{a b c}  \times   
	 g^{ij} (k^\mu_2 - k^\mu_1) 
	 ,\\
	i \frac{\delta^4 S}{\delta W^{a}_{i} 
			    \delta W^{b}_{j} 
			    \delta G^{c}_{\mu} 
			    \delta G^{d}_{\nu} 
			    } 
& = & 
			    -i g_s^2 
			    \left( 
			    f^{a c e } f^{e b d} 
			    + f^{a d e } f^{ e b c  } 
			    \right) \times 
	g^{ij} g^{\mu\nu}, \\
	i \frac{\delta^4 S}{\delta W^{a}_{i} 
			    \delta W^{b}_{j} 
			    \delta W^{c}_{k} 
			    \delta W^{d}_{l} 
 			    } 
& = & 
	- i 
	  g_s^2
	\left(
		 f^{a c e } f^{e b d} 
		+
		f^{a d e } f^{e b c} 
		\right)  \times
	  g^{i j}
	  g^{k l}
		\nonumber\\
&  & - 
	i g_s^2
	\left(
		f^{a b  e} f^{e c d} 
		+
		f^{c b  e}  f^{e a d} 	
	\right)  \times 
	  g^{i k}
	  g^{j l}
		\nonumber\\
&  & -  
	i g_s^2 
	\left(
		f^{a b e } f^{e d c } 
		+
		 f^{d b e} f^{e a c} 
		\right) \times 
	  g^{i l}
	  g^{j k},
\end{eqnarray}
	where an overall momentum conservation delta-function is implied.

	In a general case, one may consider the following  
	form of the $\varepsilon$-scalar four-vertex\cite{Harlander:2006rj}:
\begin{equation}
	{\mathcal L}_{\varepsilon\varepsilon\varepsilon\varepsilon} 
	= - \frac{1}{4} \sum\limits_{r=1}^R \lambda_r H^{abcd}_r 
	W_i^a W_j^c W_i^b W_j^d
	\label{e-scalar:4vertex}
\end{equation}
	Clearly, tensors $H$ are symmetric under permutations 
	$a - b$, $c-d$ and
	$(a,b)-(c,d)$.  
	For the gauge group $SU(N)$ the dimensionality $R$ of the basis 
	of rank-four tensors $H^{abcd}_r$ that
	are symmetric with respect to $(a,b)$ and $(c,d)$ 
	exchange is given by $R=2$ for $SU(2)$,
	$R=3$ for $SU(3)$ and $R=4$ for $SU(N), N\geq4$.

	The Feynman rule for the vertex \eqref{e-scalar:4vertex} 
	with external $\varepsilon$-scalars reads 
\begin{equation}
i \frac{\delta^4 S}{\delta W^a_i \delta W^b_j \delta W^c_k \delta W^d_l}
	 =  
		- i 2 \sum\limits_{r=1}^R \lambda_r
		\left( g^{ij} g^{kl} H^{abcd}_r 
		    +  g^{ik} g^{jl} H^{acbd}_r 
		    +  g^{il} g^{jk} H^{adbc}_r \right)
		\label{feynman:4-vertex}
\end{equation}

	One can choose $H^{abcd}_r$ to be
\begin{subequations}
	\label{Habcd}
\begin{eqnarray}
	H_1^{abcd} & = & \frac{1}{2}\left(f^{ace} f^{bde} + f^{ade} f^{bce} \right) 
	\label{Habcd:1} \\
	H_2^{abcd} & = & \delta^{ab} \delta^{cd} + \left( \delta^{ac} \delta^{bd} + \delta^{ad} \delta^{bc}\right) 
	\label{Habcd:2} \\
	H_3^{abcd} & = & \frac{1}{2} \left( \delta^{ac} \delta^{bd} + \delta^{ad} \delta^{bc}\right)  
	- \delta^{ab} \delta^{cd}  
	\label{Habcd:3}
\end{eqnarray}
\end{subequations}
	If the QCD is embedded in a model with (softly broken) supersymmetry,
	$\varepsilon$-scalar four-vertex is related by
	(double) supersymmetry transformation to the corresponding
	gluon vertex. Consequently, if the symmetry is not explicitly
	broken by regularization and renormalization, couplings
	for gluon and $\varepsilon$-scalar four-vertices are renormalized 
	in the same way, i.e., 
	$\lambda_1 = g_s^2, \lambda_i = 0, i>1$. 

	We proceed with the fermion sector of the QCD. 
	The interaction Lagrangian
	in four dimensions looks like
\begin{equation} \label{QCD:fermion-gluon}
	\delta {\cal L} = - g_s \overline q \gamma^\mu G^a_{\mu} T^a q,
\end{equation}
	where $T^a$ is a generator of $SU(3)$ in fundamental representation.
	After dimensional reduction \eqref{QCD:fermion-gluon}
	induces interaction of 
	the $\varepsilon$-scalars with quarks, i.e.,
\begin{equation} \label{lag:escalar-quark}
	\delta {\cal L}_\varepsilon 
	= - g_s \overline q \gamma^i W^a_i T^a q
\end{equation}
	It should be noticed that gamma matrices $\gamma_i$
	with index $i$ from $2 \varepsilon$-subspace, anticommute with
	``ordinary'' gamma-matrices that represent a vector with respect
	to $d$-dimensional Lorentz group. Another property
	is that the product of two identical gamma matrices $\gamma_i$ is
	equal to $-1$. All these properties clearly come from the
	relations
\begin{eqnarray}
	\{ \gamma_\mu, \gamma_\nu \} & = & 2 g_{\mu\nu} \\
	\{ \gamma_\mu, \gamma_i \} & = & 0 \\
	\{ \gamma_i, \gamma_j \} & = & 2 g_{ij}
\end{eqnarray}

	Again the term \eqref{lag:escalar-quark}
	alone is invariant under gauge transformations, 
	so the renormalization of the corresponding coupling may 
	not coincide with that of $g_s$. Consequently,
	we rewrite \eqref{lag:escalar-quark} in the following way:
\begin{equation} \label{lag:escalar-quark-yukawa}
	\delta {\cal L}_\varepsilon 
	= - g_y \overline q \gamma^i W^a_i T^a q,
\end{equation}
	where $g_y$ denotes evanescent Yukawa coupling\cite{Jack:1993ws}.  It can be 
	set to be equal to $g_s$ at any scale. However, 
	one should be careful trying to make a prediction
	at a different scale due to different running of evanescent and real couplings.
	A Feynman rule for \eqref{lag:escalar-quark-yukawa} reads
\begin{equation} \label{feynman:escalar-quarks}
	i \frac{\delta^3 S}
	{\delta_L \, \overline q^{k_1}_{\alpha_1} \,
	 \delta_R \, q^{k_2}_{\alpha_2} \,
	 \delta \, W^{a}_{i}} 
	 = - i g_y T_{k_1 k_2}^a \times \gamma^i_{\alpha_1 \alpha_2},
\end{equation}
	where $\delta_{L(R)}$ denotes left (right) functional derivative,
	$(k_1, k_2)$ are color indices and $(\alpha_1, \alpha_2)$ 
	correspond to Dirac spinor indices.
	In general, one should distinguish $\gamma_i$ from 
	$\gamma^i = g^{ij} \gamma_i = - \gamma_i$.
	However, in almost all practical calculations 
	one ``scalarize'' the expression for a Feynman amplitude
	by contraction of its free indices with an appropriate projector.
	In a scalarized expression the relevant property 
	is $g^{ij} \,g_{ij} = 2 \varepsilon$. 

	Finally, there are gauge fixing and ghost terms in the Lagrangian
	of four-dimensional QCD
\begin{equation}
	\delta \mathcal{L}  =  
	  -   \frac{1}{2 \xi} \left( \partial_\mu G_a^\mu \right)^2
	- \overline{c}^a \partial^{\mu} \left( \partial_\mu \delta^{ab} + g f^{abc} G^c_{\mu} \right) c^b. 
\end{equation}
	Clearly, after dimensional reduction these terms do not contribute
	to the interaction Lagrangian for the $\varepsilon$-scalars.
\section{The $\varepsilon$-scalars in SUSY QCD}\label{sec:escalar-sqcd}
	
	In SUSY QCD $\varepsilon$-scalars interact not only with
	quarks and gluons but also with squarks and gluinos.
	Actually, it is $\varepsilon$-scalars that balance the number
	of fermionic and bosonic degrees of freedom in the
	$d$-dimensional world.

	As in the previous section, consider the four-dimensional
	form of the relevant part of the SUSY QCD Lagrangian 
\begin{eqnarray}\label{lag:SUSYQCD}
	\delta \mathcal{L} & = & \frac{i}{2} 
	\bar{\tilde g}^a D^{ab}_{\mu} \gamma^{\mu} \tilde g^b 
	+ \sum\limits_{n=1,2}\left(D_{\mu} \tilde q_n\right)^* \, 
		              \left(D^{\mu} \tilde q_n\right).
\end{eqnarray}
	Here $D^{ab}_{\mu}$ is a covariant derivative 
	for gluinos $\tilde g$ (see \eqref{derivative:adjoint})
	and $D_{\mu}$ denotes a covariant derivative for squarks $\tilde q_n$ 
	that belong to fundamental representation of the color group. 
	Notice that in \eqref{lag:SUSYQCD} we do not 
	write explicitly summation over quark flavours.

	After dimensional reduction some of gluon fields 
	that enter into covariant derivatives in \eqref{lag:SUSYQCD} become
	$\varepsilon$-scalars. 
	Therefore, gluino interaction with 
	$\varepsilon$-scalars reads
\begin{eqnarray}\label{lag:escalar-gluino}
	\delta \mathcal{L}_{\varepsilon} & = &  	
	i \frac{g_s}{2} f^{abc} 
	\bar{\tilde g}^a \gamma^{i} \tilde g^b \, W_i^c 
\end{eqnarray}
	and the Feynman rule is
\begin{equation}
\label{feynman:escalar-gluino}
	i \frac{\delta^3 S}
	{\delta_L \, \bar{\tilde g}^{a_1}_{\alpha_1} \,
	 \delta_R \, \tilde g^{a_2}_{\alpha_2} \,
	 \delta \, W^{a_3}_{i}} 
	 = - g_s f^{a_1 a_2 a_3} \times \gamma^i_{\alpha_1 \alpha_2}.
\end{equation}
In \eqref{feynman:escalar-gluino} the factor $1/2$ from \eqref{lag:escalar-gluino} is canceled due to a majorana nature of gluino.
	For the squark-$\varepsilon$-scalar interactions we have only 
	four-vertices. Three-vertices inevitably involve derivatives
	with respect to the coordinates that belong to $2\varepsilon$
	subspace and, therefore, vanish. 
	Accordingly,
\begin{equation}\label{lag:escalar-squarks}
	\delta \mathcal{L}_{\varepsilon} =
	g^{ij} g_s^2 
	\sum\limits_{n=1,2} \tilde q^*_n T^a T^b \tilde q_n W^a_i W^b_j 
\end{equation}
	and the Feynman rule is
\begin{equation} \label{feynman:escalar-squarks}
i \frac{\delta^4 S}{\delta \, (\tilde q^*)^{l_1}_{n_1} \,
	\delta \, \tilde q^{l_2}_{n_2} 
	\delta W^{a_1}_{i_1} 
	\delta W^{a_2}_{i_2} }
	 =  
	 i g_s^2 \left(T^{a_1} T^{a_2} + T^{a_2} T^{a_1}\right)_{l_1 l_2}
	 	\delta_{n_1 n_2} \times g^{ij}.
\end{equation}
	Here again $(l_1,l_2)$ are color indices and $(n_1\, n_2)$
	numerate different squarks from \eqref{lag:SUSYQCD}. 
	Generalization to the multi-flavour case is straightforward. 
	Since strong interactions are flavour-blind, 
	the ``generalization'' amounts to additional 
	``flavour'' Kronecker delta.

	All needed Feynman rules are summarized in Table~\ref{fig:feynman_rules}.
\begin{table}[t]
\tbl{Feynman rules for gluon $\varepsilon$-scalars. All momenta are incoming.  
	Metric tensor $g^{ij}$ corresponds to 
	$2\varepsilon$-dimensional space, $g^{ii} = -1$ (no summation). Gamma matrices $\gamma^i$ 
	carry  $2\varepsilon$-dimensional indices.   
	Generators and structure constants of
	$SU(3)$ are denoted by $T^a$ and $f^{abc}$, respectively. 
	For the definition of $H_r^{abcd}$ see \eqref{Habcd}.
	}{
	\begin{tabular}{cc}
\includegraphics[scale=0.6]{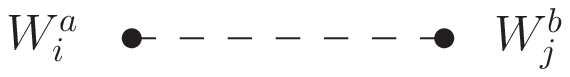} &
\figbox{		
	$ - i g^{ij} \, \delta^{ab} \, \left(k^2 - \meps^2\right)^{-1} $
}\\
\includegraphics[scale=0.6]{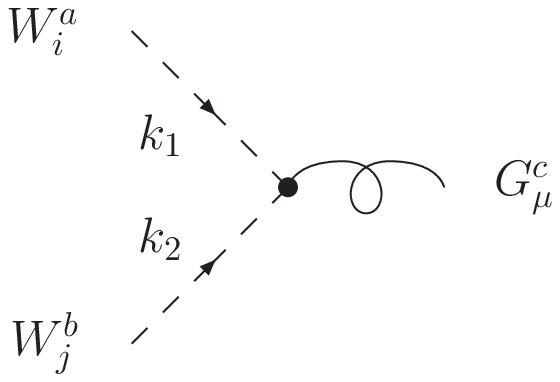} & 
\figbox{		
	$ g_s f^{abc} \times g^{ij} (k^\mu_2 - k^\mu_1) $  
}\\
\includegraphics[scale=0.6]{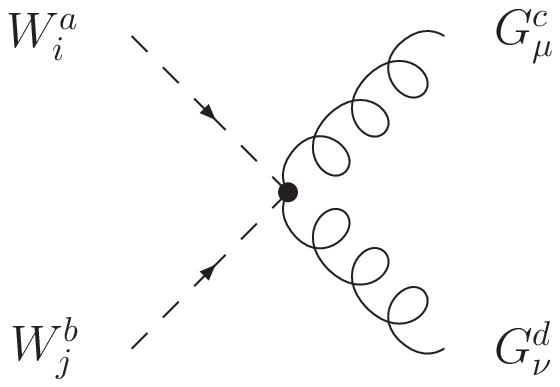} &
\figbox{
	$ -i g^2_s \left (f^{ace} f^{ebd} + f^{ade} f^{ebc} 
		\right) \times g^{\mu\nu} g^{ij} $ 
	} \\
\includegraphics[scale=0.6]{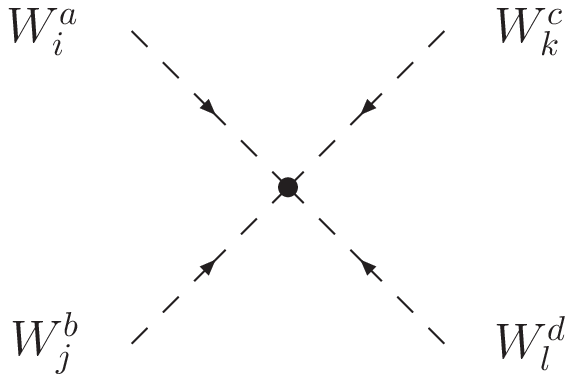} 
		& 
\figbox{
		$- i 2 \sum\limits_{r=1}^3 \lambda_r
		\left( g^{ij} g^{kl} H^{abcd}_r 
		    +  g^{ik} g^{jl} H^{acbd}_r +  g^{il} g^{jk} H^{adbc}_r \right) $
	} \\
\includegraphics[scale=0.6]{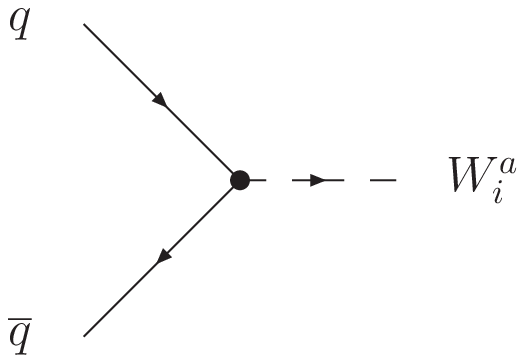} 
		& 
\figbox{
	$ - i g_y T^a \times \gamma^{i}$
	} \\
\includegraphics[scale=0.6]{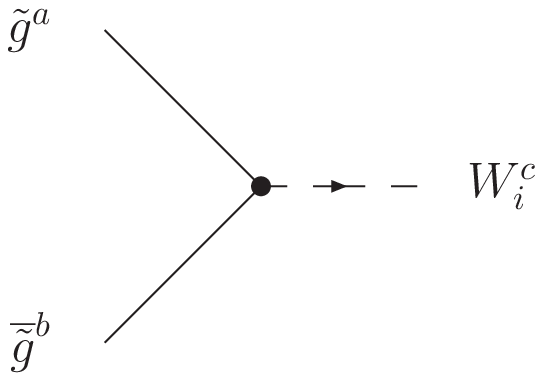} 
	& 
\figbox{
	$ - g_s f^{abc} \times \gamma^{i}$
	}  \\
\includegraphics[scale=0.6]{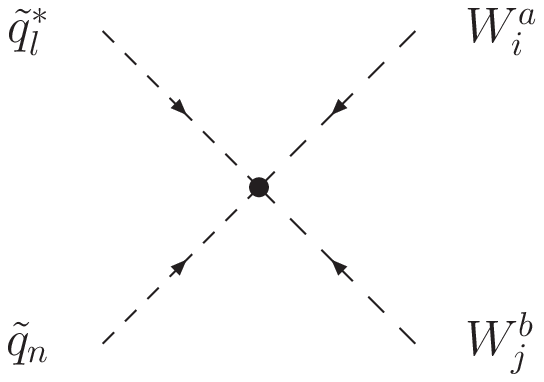} 
	& 
\figbox{
	$  g_s^2 \left(T^a T^b + T^b T^a\right) \delta_{ln} \times g^{ij}$
	}  
\end{tabular} \label{fig:feynman_rules} 
}
\end{table}

\section{FeynArts implementation of the $\varepsilon$-scalar Lagrangian}
	The FeynArts package allows one to generate
	needed diagrams and Feynman amplitudes automatically. 
	The MSSM has already been implemented in FeynArts (see Ref.~\refcite{Hahn:2001rv}). 
	The model information is contained in two special files:
	The \emph{generic model file} defines representation of the kinematical
	quantities. The \emph{classes model file} sets up the particle
	content and specifies the actual couplings.
 
	One of the crucial properties of $\varepsilon$-scalars is
	that they carry $2 \varepsilon$-dimensional indices
	(one may say that we have $2 \varepsilon$ scalars).
	This property fixes ``kinematical'' structure of
	$\varepsilon$-scalar vertices, i.e., possible
	products of $g^{ij}$ and other Lorentz objects
	which can appear in a vertex. Moreover, it does not depend
	on the group to which the $\varepsilon$-scalars belong.
	Consequently, the property can be realized at the \emph{generic} level.  
	For this purpose we write an addendum {\tt LorentzEps.gen} for 
	the generic model file {\tt Lorentz.gen}.  

	The kinematical structure of vertices with $\varepsilon$-scalars
	is more like than of gauge bosons that of ordinary scalars.
	Instead of using a pre-defined generic scalar field {\tt S}, 
	it is convenient to
	introduce a new generic field $W$ ({\tt SE} in FeynArts). 
	The field $W$ represents \emph{generic} 
	$\varepsilon$-scalars and carries
	new kinematic index $i = \mathtt{Index[Escalar]}$.
	For the field $W_i$ we assume that there is no external 
	wave function and a  propagator has the form:
\begin{eqnarray} \label{feynman:escalar-prop}
	\langle W_i(-k) | W_j(k) \rangle = 
	- i \frac{g^{ij}}{k^2 - \meps^2}.
\end{eqnarray}	
	The mass $\meps^2$ for the scalars is introduced
	due to the fact that there is no symmetry that keeps 
	$\varepsilon$-scalars massless at each order of perturbation
	theory. 

	Let us summarize the generic kinematical structure of the 
	couplings. We use the same notation as in Ref.~\refcite{FeynArts}
\begin{eqnarray}
C(W_{i}, W_{j}, W_{k}, W_{l}) & = & 
	\vec{G}_{WWWW} \, \cdot \,
	\begin{pmatrix}
		g^{i j} g^{k l} \\
		g^{i k} g^{j l} \\
		g^{i l} g^{j k} 
	\end{pmatrix}_{+} \label{fa-generic:e-e-e-e} \\
C(W_{i}, W_{j}, V_{\mu}, V_{\nu}) & = & 
	G_{WWVV} \, \cdot \,
	\left( g^{i j} g^{\mu \nu} \right)_+ 
	\label{fa-generic:e-e-v-v}\\
C(W_{i}(k_1), W_{j}(k_2), V_{\mu}(k_3)) & = & 
	G_{WWV} \, \cdot \,
	\left( g^{i j} (k_2^{\mu}-k_1^{\mu})\right)_-
	\label{fa-generic:e-e-v}\\
C(W_{i}, W_{j}, S, S) & = & 
	G_{WWSS} \, \cdot \,
	\left( g^{i j}\right)_+
	\label{fa-generic:e-e-s-s}\\
C(F, F, W_{i}) & = & 
	\vec{G}_{FFW} \, \cdot \,
	\begin{pmatrix}
		\gamma^i \omega_- \\
		\gamma^i \omega_+
	\end{pmatrix}_{-} \label{fa-generic:f-f-e} \\
C(W_i(k_1), W_j(k_2)) & = & 
	\vec{G}_{WW} \, \cdot \,
	\begin{pmatrix}
		g^{ij} (k_1 k_2) \\
		g^{ij}
	\end{pmatrix}_{+}. \label{fa-generic:e-e} 
\end{eqnarray}
	Here antisymmetric couplings are labeled by a subscript $-$ and
	symmetric ones by a subscript $+$. The fields $V_{\mu}, W_i, F, S$ 
	correspond to generic vector, $\varepsilon$-scalar, fermion 
	and ordinary scalar fields. Actual coupling vectors $\vec{G}$ 
	should be defined for each particular model.
	
	Note that for the metric tensor $g^{ij}$ and for
	the Dirac matrices $\gamma^i$ no new symbols were defined. 
	We use the following notation:
\begin{eqnarray*}
	\gamma^i & = & \mathtt{DiracGamma\left[Index[Escalar,i]\right]}, \\
	g^{ij} & = & \mathtt{MetricTensor[Index[Escalar,i],Index[Escalar,j]]}.
\end{eqnarray*}

	To implement gluon $\varepsilon$-scalars in the 
	context of (SUSY) QCD 
	a new classes model file is written {\tt ESCALAR.mod}.	
	Actually, the file only extends particle content and
	adds new couplings to the MSSM model {\tt MSSMQCD.mod}.
	The generic (nonsuperymmetric)
	structure of vertices described in~\ref{sec:escalar-qcd}
	is implemented.	This allows one to use the same model file for 
	the QCD and SUSY QCD. Actual coupling vectors for these models
	can be easily inferred from the expressions given above\footnote{
	See \url{http://theor.jinr.ru/~bednya/pmwiki/pmwiki.php?n=Main.Escalars}}.

\providecommand{\href}[2]{#2}\begingroup\raggedright\endgroup

\end{document}